\documentclass[
aps,
pre,
superscriptaddress,
twocolumn,
showpacs,
showkeys,
]{revtex4-1}

\usepackage{graphicx}
\usepackage{dcolumn}
\usepackage{bm}
\usepackage{subfigure}
\usepackage{amsthm}

\begin{document}
\preprint{Physical Review E}

\title{Local structure entropy of complex networks}


\author{Qi Zhang}
\affiliation{School of Computer and Information Science, Southwest University, Chongqing 400715, China}
\author{Meizhu Li}
\affiliation{School of Computer and Information Science, Southwest University, Chongqing 400715, China}
\author{Yuxian Du}
\affiliation{School of Computer and Information Science, Southwest University, Chongqing 400715, China}
\author{Yong Deng}
\email{ydeng@swu.edu.cn; prof.deng@hotmail.com}
\affiliation{School of Computer and Information Science, Southwest University, Chongqing 400715, China}%
\affiliation{School of Automation, Northwestern Polytechnical University, Xian, Shaanxi 710072, China}



\date{\today}

\begin{abstract}
    Identifying influential nodes in the complex networks is of theoretical and practical significance. There are many methods are proposed to identify the influential nodes in the complex networks. In this paper, a local structure entropy which is based on the degree centrality and the statistical mechanics is proposed to identifying the influential nodes in the complex network.
    In the definition of the local structure entropy, each node has a local network, the local structure entropy of each node is equal to the structure entropy of the local network. The main idea in the local structure entropy is try to use the influence of the local network to replace the node's influence on the whole network.
    The influential nodes which are identified by the local structure entropy are the intermediate nodes in the network. The intermediate nodes which connect those nodes with a big value of degree.
    We use the  $Susceptible-Infective$ (SI) model to evaluate the performance of the influential nodes which are identified by the local structure entropy. In the SI model the nodes use as the source of infection. According to the SI model, the bigger the percentage of the infective nodes in the network the important the node to the whole networks. The simulation on four real networks show that the proposed method is efficacious and rationality to identify the influential nodes in the complex networks.

\end{abstract}



\maketitle

\section{Introduction}
The complex networks is a system which composed of many interacting parts \cite{kim2008complex,newman2003structure}. There are many real systems can be modeled as the complex networks. Identifying influential nodes in the complex networks is of theoretical and practical significance. There are many methods have proposed to identify influential nodes, such as the degree centrality, the betweenness centrality and so on \cite{chen2012identifying,freeman1977set,anand2009entropy}.

The degree centrality method is very simple but can not illuminate the global characteristic. The betweenness centrality method is useful but can not been implemented in the big scales complex networks. In order to find a more reasonable method to identifying influential nodes in the big scales complex networks, we proposed a new method which has merged the statistical mechanic and the degree centrality of the complex networks which is named local structure entropy of the complex networks .

The local structure entropy is defined based on the shannon entropy and the degree centrality. The results of our test show that the proposed method is useful and efficient.

The rest of this paper is organised as follows. Section \ref{Rreparatorywork} introduces some preliminaries of this work, such as the definition of the shannon entropy, the degree centrality and the betweenness centrality. In section \ref{new}, a new method to identifying influential nodes in the complex networks is proposed. The application of the proposed method is illustrated in section \ref{application}. Conclusion is given in Section \ref{conclusion}.

\section{Preliminaries}
\label{Rreparatorywork}
In this section, we introduce some core concepts which will be used in this paper.
\subsection{Degree centrality}
\label{Degree}
\begin{figure}
  \centering
  \includegraphics[scale=4]{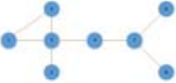}\\
  \caption{In the network, depends on the degree the nodes can been classified into different set. In the set $S_1={2,8,6}$ each node's degree is equal to 1. In the set $S_2={1,5,4}$ each node's degree is equal to 2. In the set $S_3={7}$ each node's degree is equal to 3. In the set $S_4={3}$ each node's degree is equal to 4. The degree is the simple and reasonable characteristic use to describe the structure property of the complex network.}
  \label{degree_example}
\end{figure}

The degree of one node in a network is the number of the edges connected to the node. Most of the complex networks' properties are based on the degree distribution, such as the clustering coefficient, the community structure and so on. In the network, ${k_i}$ represents the degree of the $i$th vertex \cite{newman2003structure}. The details of the degree centrality are shown in the Fig \ref{degree_example}.

\subsection{Betweenness centrality}
\label{betweenness centrality}
The betweenness centrality is an important index which can be used to illuminate the importance of the nodes. It is defined based on the shortest path of the network \cite{barthelemy2004betweenness}.

\begin{figure}
  \centering
  \includegraphics[scale=3]{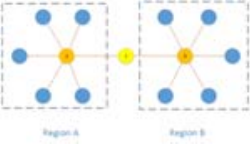}\\
  \caption{The vertex 1 in the figure connect the region A and region B. The degree of the vertex 1 is 2, but all shortest path from the nodes in region A to the nodes in region B have to go to through the vertex 1. The vertex 1 have a small value of degree but a large value of betweenness. In fact, vertex 1 is here a cut-vertex; its removal will break the network into two disconnected components.}
  \label{betweenness}
\end{figure}

The betweenness centrality of the complex networks is defined as follows \cite{barthelemy2004betweenness}:

\begin{equation}\label{Betweenness}
bet(i) = \frac{{\upsilon (i)}}{{\sum {{\sigma _{st}}} }}(s \ne i \ne t)
\end{equation}

In the Eq. (\ref{Betweenness}), the ${{\sigma _{st}}}$ is the number of the shortest path from vertex $s$ to vertex $t$, ${\upsilon (i)}$ is the number of the shortest path which have go to through the vertex $i$ \cite{barthelemy2004betweenness}. The details of the betweenness centrality is shown in the Fig.\ref{betweenness}. However, it is clear that the betweenness centrality is hard to calculate when the network has a big scale.
\subsection{Shannon entropy}
\label{shannon entropy}

The shannon entropy which is named as the information entropy \cite{shannon2001mathematical} is the basic of the information theory. In the information theory, the shannon entropy is used to measure the unpredictability of the information content. In other words, the shannon entropy is used to measure the uncertainty in the system which is described by the probability theory.

The classical form of the Shannon entropy is defined as follows \cite{shannon2001mathematical}.

\begin{equation}\label{Shannon-entropy}
 {E_{Shannon}}{\rm{ = }}\sum\limits_{i = 1}^n {{p_i}} \log ({p_i})
\end{equation}
\subsection{Susceptible and infective model in the network}
In order to improve the influential of the nodes. We use the SI model (Susceptible and infective model) in the network. The process of the infection can be divide into three steps:

\textbf{Step 1}: Choosing one of the node as the source of infection. Set the times($Infect-T$) of the infective node to infect other normal nodes.

\textbf{Step 2}: Find the neighbour node of the infection source node. Infecting the neighbour nodes randomly of an probability. Repeating this process in $Infect-T$ times.

\textbf{Step 3}: After $Infect-T$ times infection, check the numbers of the infective nodes in the network. Calculating the probability of infection.

The probability of the infection represents the influence of the nodes in the network. The value of the infective times in the process is decided by the scale of the network. The large the scale of the target network, the big the value of infective times.
\section{Some problems in the existing methods}
\begin{figure}
  \centering
  \subfigure[Network A]{
    \label{Local_detail:subfig:a} 
    \centering
    \includegraphics[scale=0.4]{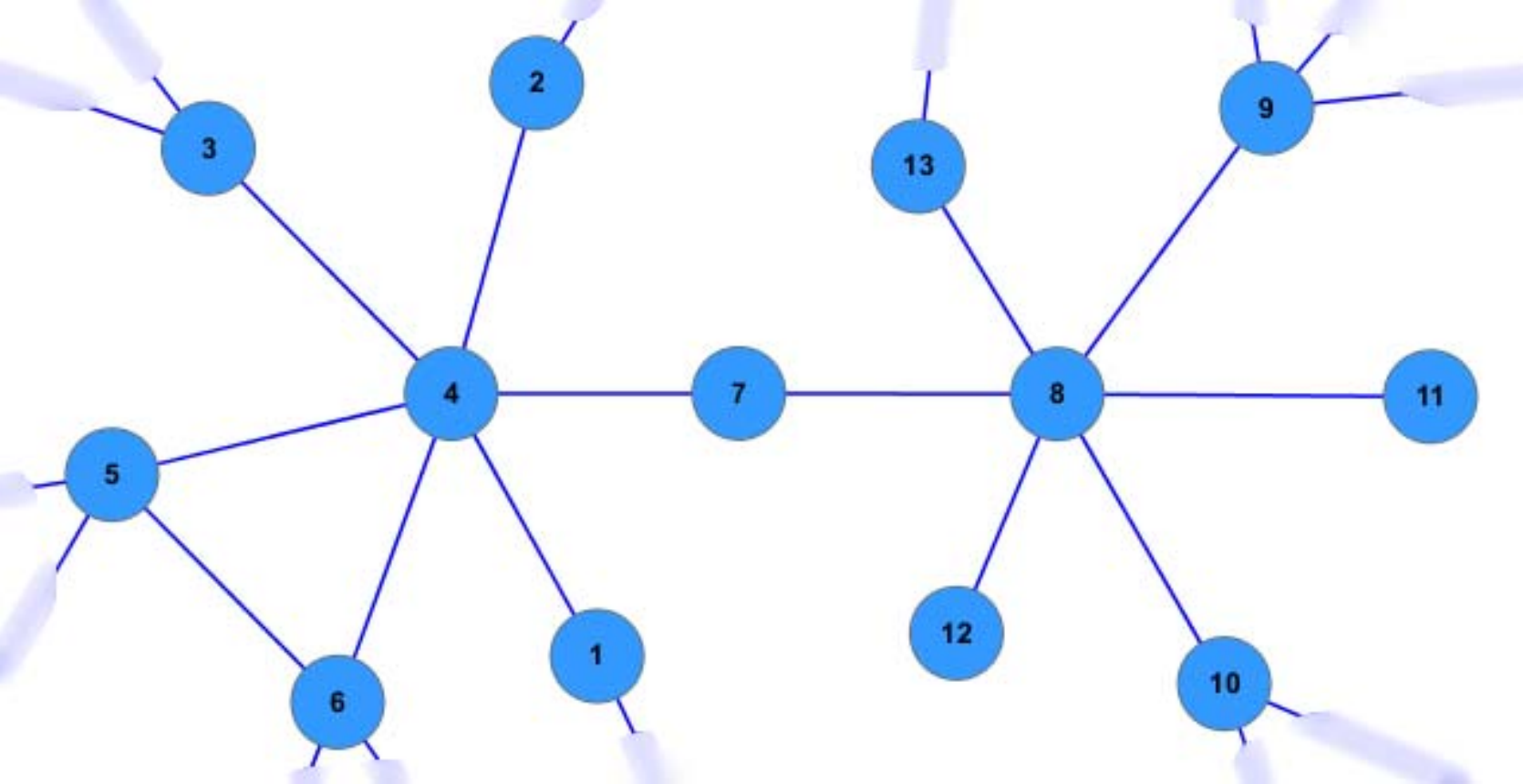}}
  \hspace{0.5cm}
    \subfigure[The detail of node 8]{
    \label{Local_detail:subfig:b} 
    \centering
    \includegraphics[scale=0.3]{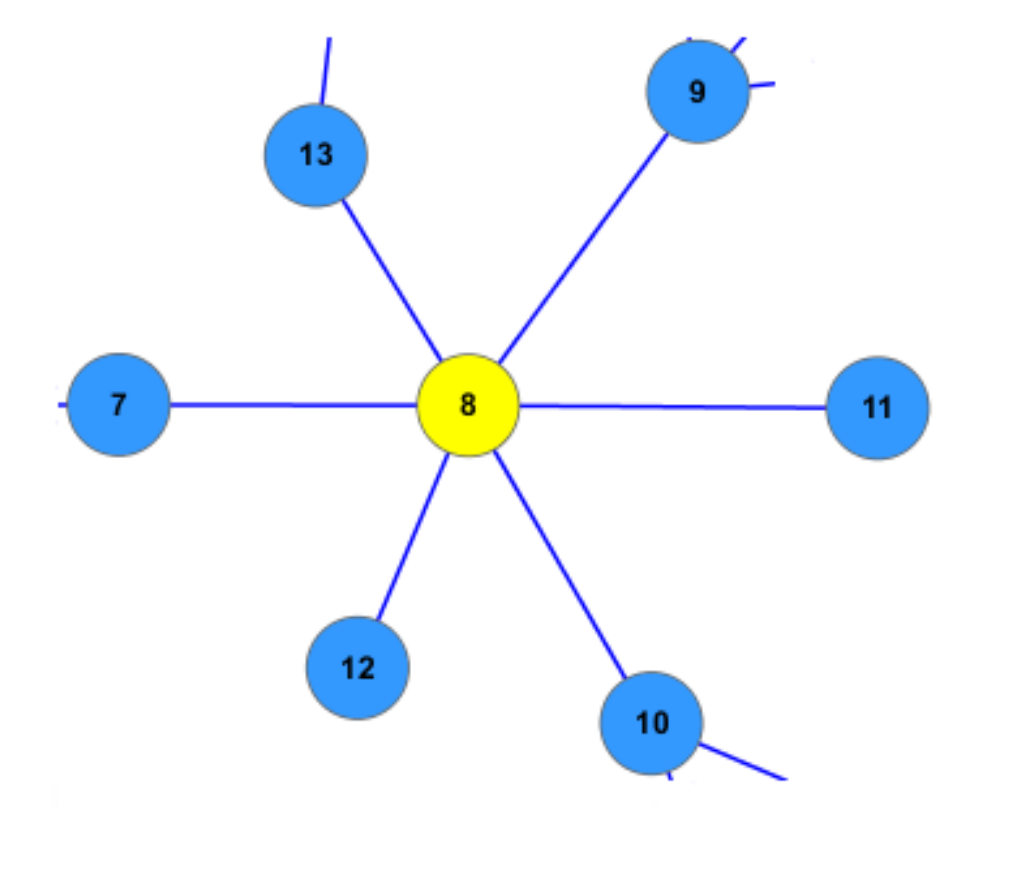}}
  \hspace{0.5cm}
    \subfigure[The detail of node 4]{
    \label{Local_detail:subfig:c} 
    \centering
    \includegraphics[scale=0.6]{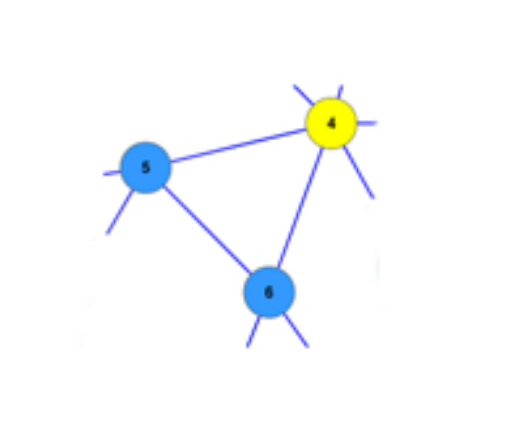}}
  \hspace{0.5cm}
      \subfigure[The detail of node 7]{
    \label{Local_detail:subfig:d} 
    \centering
    \includegraphics[scale=0.3]{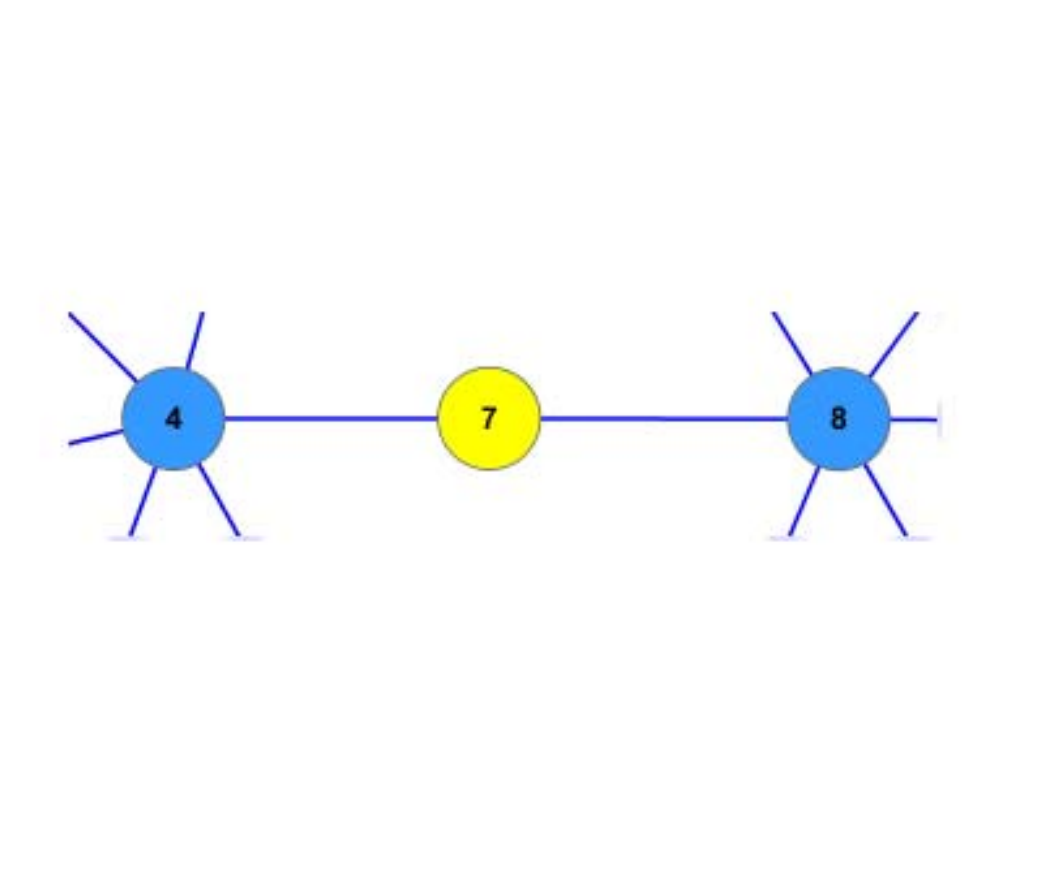}}
  \hspace{0.5cm}
        \subfigure[The detail of node 11]{
    \label{Local_detail:subfig:e} 
    \centering
    \includegraphics[scale=0.3]{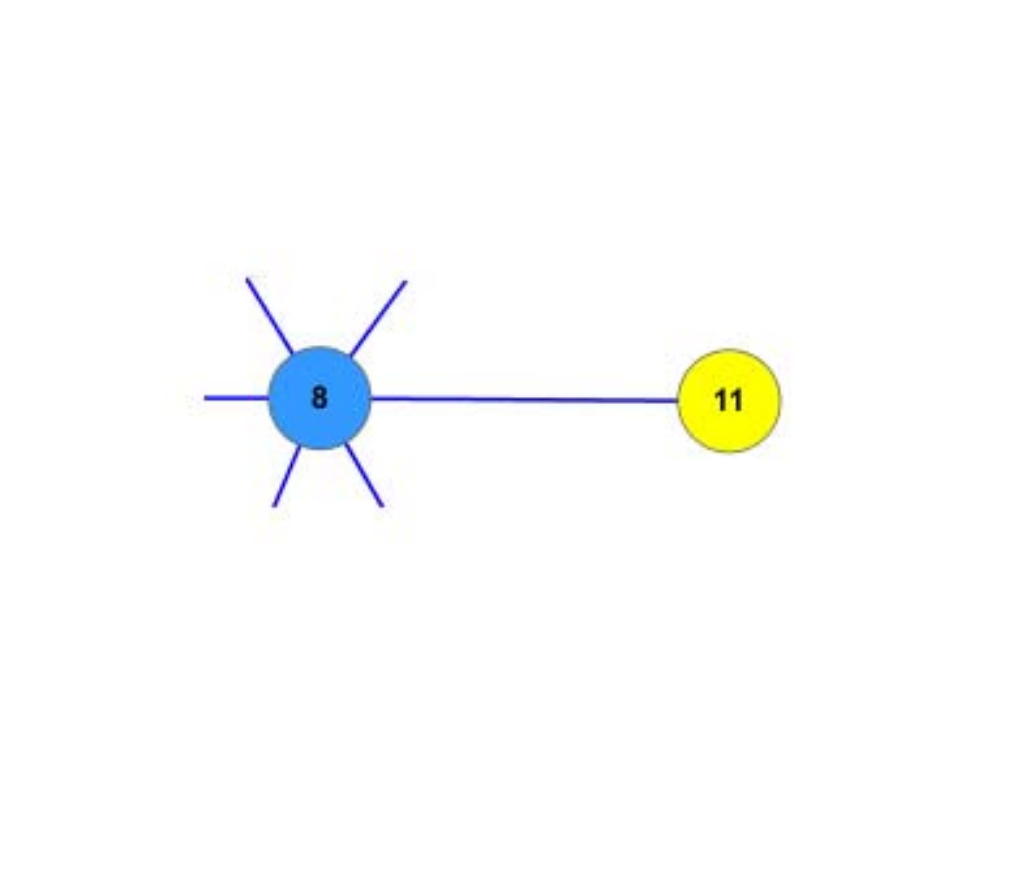}}
  \caption{The Network A shows in the subfigure a is a local part of a complex network. Other subfigures have shown some special nodes.In this figure, there are 5 subfigures. A part of one large scale network is shown in the subfigure (a). The detail of node 8 is shown in the subfigure (b). The node 8 has a big value of degree. In the local network around node 8,depends on the degree centrality, the node 8 is the veritably central node. In the subfigure (c), the 3 nodes have an equality relationship to each other. In the subfigure (d), the node 7 has a small value of degree. In other words, depends on the degree centrality the node 7 has a small influence to the whole network. However, from the global view the node 7 connect two nodes which have a big value of degree, it mean the node 7 is important for the whole network. The degree centrality can not be used to describe the nodes which have the same structure property of nodes 7 and the betweenneess centrality can not be used in the large scale network. It need a new method to describe those nodes which have the same structure property of nodes 7.
}\label{Local_detail}
\end{figure}

There are many methods can be used to identifying the influential nodes in the complex networks, such as the degree centrality, the betweenness centrality and so on \cite{PhysRevE.90.032812,PhysRevE.84.046101,PhysRevE.82.046117}. However, in the large scale networks to calculate the betweenness centrality is inefficiently. The degree centrality is lack of information to describe the special nodes' structure property in the large scale network. The details are shown in the Fig. \ref{Local_detail}.

Based on the existing researches, the degree centrality is a effective method to identifying the influential nodes in the complex network. However, it is an incomplete method. The degree centrality of the nodes only have considered the direct connection to the target node. There are so many nodes in the complex networks which is in the same structure property of the node 7 in the Fig.\ref{Local_detail:subfig:d} \cite{PhysRevE.70.026109,PhysRevE.80.046114}. This kind of nodes have a small value of degree, but a big influence in the local network. In order to describe the structure of the complex network more effective and more convenient, we proposed a new method based on the degree centrality and the shannon entropy.

\section{Local structure entropy of the complex networks}
\label{new}
Because of the effective and convenient of the degree centrality \cite{Gao20135490,Gao2014130,Wei20132564,Du201457,Li201447,Ji201487}, the new method proposed in this paper is based on the degree centrality too. However, in the new method, the influence of the neighbour nodes of the target node is considered.

The main idea of the definition of the new method is that the influence of the target node's neighbour is contained. Therefore, a local network around each node is established by us. The influence of the node on the whole network is replaced by the influence of the local network on the whole network.

There are many researches of complex network are based on the statistical mechanics \cite{albert2002statistical,tsallis1988possible,tsallis2009nonadditiveBJP}, such as the information dimension \cite{Zhang2015707,wei2014informationdimension}, the structure entropy \cite{anand2009entropy}. The researches show that the statistical mechanics is an useful method to describe the structure property of the complex networks.
The structure entropy of a network is used to describe the structure complexity of it. The more the complex of the network, the big the value of the structure entropy. Depends on this definition of the structure entropy, the influence of each node can be described by the local network's structure entropy.

Depends on the definition of the local network, the degree centrality and the shannon entropy, a local structure entropy of the complex network is proposed in this paper to identifying the influential nodes in the complex network.

The definition of the local structure entropy can be separate into three steps:

\textbf{Step 1 Creating a local network}: Choosing one of the node in the network as a central node of the local network. The neighbour nodes of the central node is contained in the local network. In other words, the local network of each node in the complex network is a part of the complex network which contains the target node and the neighbour nodes of it.

\textbf{Step 2 Calculating the unit of the local structure entropy}: Calculating the degree of each node in the local network and the total number of the degree in the local network. The unit of the local structure entropy can be represents as the ${p_{ij}}$, it is defined in the Eq.(\ref{P_ij}).

\textbf{Step 3 Calculating the local structure entropy of each node}: The definition of the local structure entropy for each node is shown in the Eq.(\ref{Local_E}).

 \begin{equation}\label{Local_E}
    {LE_i} = \sum\limits_{j = 1}^n {{p_{ij}}} \log {p_{ij}}
 \end{equation}

Where the ${LE_i}$ represents the local structure entropy of the $i$th node in the complex networks. The $n$ is the total number of the nodes in the local network. The ${p_{ij}}$ represents the percent of degree for the $j$th node in the local network. The definition of the ${p_{ij}}$ is shown in the Eq.(\ref{P_ij}).

\begin{equation}\label{P_ij}
{p_{ij}} = \frac{{\ degree(j)}}{{\sum\limits_{j = 1}^n {\ degree(j)} }}
\end{equation}

\begin{figure}
  \centering
  \includegraphics[scale=0.5]{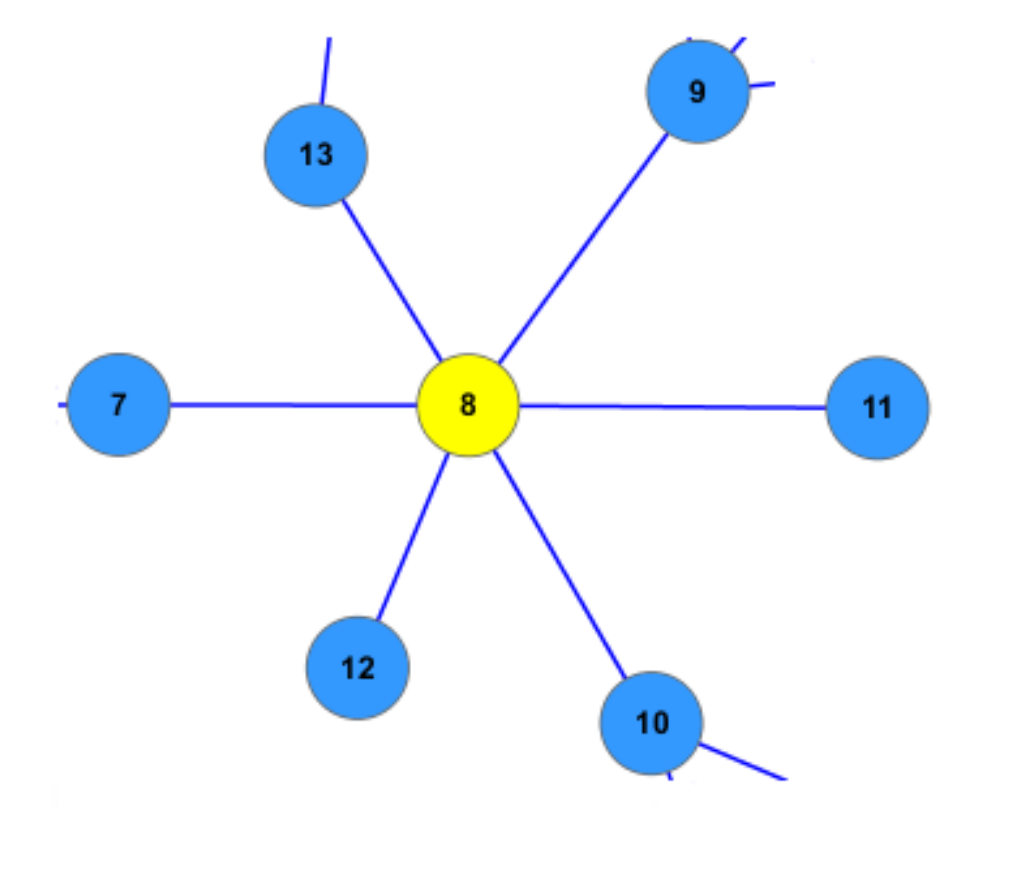}\\
  \caption{This figure is a part of the Network A shows in the Fig.\ref{Local_detail}. The central node in this local network is the node 8. The local network contains 7 nodes. The degree of the nodes in the local network is different to each other. The total number of the degree in the local network is equal to 19. The degree of node 7 is 2, the degree of node 13 is 2, the degree of node 9 is 4, the degree of node 11 is 1, the degree of node 10 is 3, the degree of node 12 is 1 and the degree of node 8 is 6. The definition of the $p_{ij}$ is shown in the set $S_8$. The $S_8 = {p_{81},p_{82},p_{83},p_{84},p_{85},p_{86},p_{87}}$. The ${p_{81}} = \frac{{\ deg ree(7)}}{{19}} = \frac{2}{{19}}$, ${p_{82}} = \frac{{\ degree(13)}}{{19}} = \frac{2}{{19}}$, ${p_{83}} = \frac{{\ degree (9)}}{{19}} = \frac{4}{{19}}$, ${p_{84}} = \frac{{\ degree(11)}}{{19}} = \frac{1}{{19}}$, ${p_{85}} = \frac{{\ degree(10)}}{{19}} = \frac{3}{{19}}$, ${p_{86}} = \frac{{\ degree(12)}}{{19}} = \frac{1}{{19}}$, ${p_{87}} = \frac{{\ degree(8)}}{{19}} = \frac{6}{{19}}$. The local structure entropy ${E_8}$ of the node 8 is equal to 1.867.}
  \label{example_local network}
\end{figure}

The detail of the process to calculate the local structure entropy is shown in the Fig.\ref{example_local network}.

In order to show the reasonable of the local structure entropy to identify the influence of the nodes in the network, a network with 21 nodes and 33 edges is used as an example to identify the most influential nodes in it. The network is named $Network B $. The value of the degree, the betweenness, the local structure entropy of each nodes in the $Network B$ are shown in the Table \ref{tab:addlabel}.

\begin{table}[htbp]
  \centering
  \caption{The influence of each node measure by different methods}
    \begin{tabular}{cccc}
    \hline
    node & degree & Betweenness & ${LE_i}$ \\
    \hline
    1     & 3     & 0.033794 & 0.695646 \\
    2     & 3     & 0.122504 & 0.814568 \\
    3     & 3     & 0.090630 & 0.772462 \\
    4     & 2     & 0.191628 & 0.619477 \\
    5     & 5     & 0.016513 & 0.940893 \\
    6     & 3     & 0.023810 & 0.662452 \\
    7     & 5     & 0.068356 & 0.911680 \\
    8     & 3     & 0.073349 & 0.814568 \\
    9     & 1     & 0.003840 & 0.366204 \\
    10    & 4     & 0.064516 & 0.761745 \\
    11    & 2     & 0.065284 & 0.657540 \\
    12    & 3     & 0.079493 & 0.851657 \\
    13    & 2     & 0.003840 & 0.664304 \\
    14    & 2     & 0.048003 & 0.643775 \\
    15    & 6     & 0.022273 & 0.977730 \\
    16    & 2     & 0.028418 & 0.619477 \\
    17    & 3     & 0.003072 & 0.808073 \\
    18    & 4     & 0.033410 & 0.814134 \\
    19    & 4     & 0.011137 & 0.814134 \\
    20    & 3     & 0.014977 & 0.797539 \\
    21    & 3     & 0.001152 & 0.745088 \\
    \hline
    \end{tabular}%
  \label{tab:addlabel}%
\end{table}%

According to the existing method and the proposed method, the most influential six nodes in the$ Network B$ are identified by those methods. The details are shown in the Table \ref{tabcompay}.
\begin{table}[htbp]
  \centering
  \caption{The most influential six nodes in the Network B}
    \begin{tabular}{lcccccc}

  \hline
    Method      & 1  & 2 & 3 & 4  & 5  & 6  \\
    \hline
  Degree        & 15 & 5 & 7 & 18 & 19 & 10 \\
  Betweenness   & 4  & 2 & 3 & 12 & 8  & 7  \\
  ${LE_i}$      & 15 & 5 & 7 & 12 & 8  & 2  \\
  \hline
    \end{tabular}
      \label{tabcompay}%
\end{table}%

\begin{figure}
  \centering
  \subfigure[The example $Network B$ ]{
    \label{Local_networkSS:subfig:b} 
    \centering
    \includegraphics[scale=0.4]{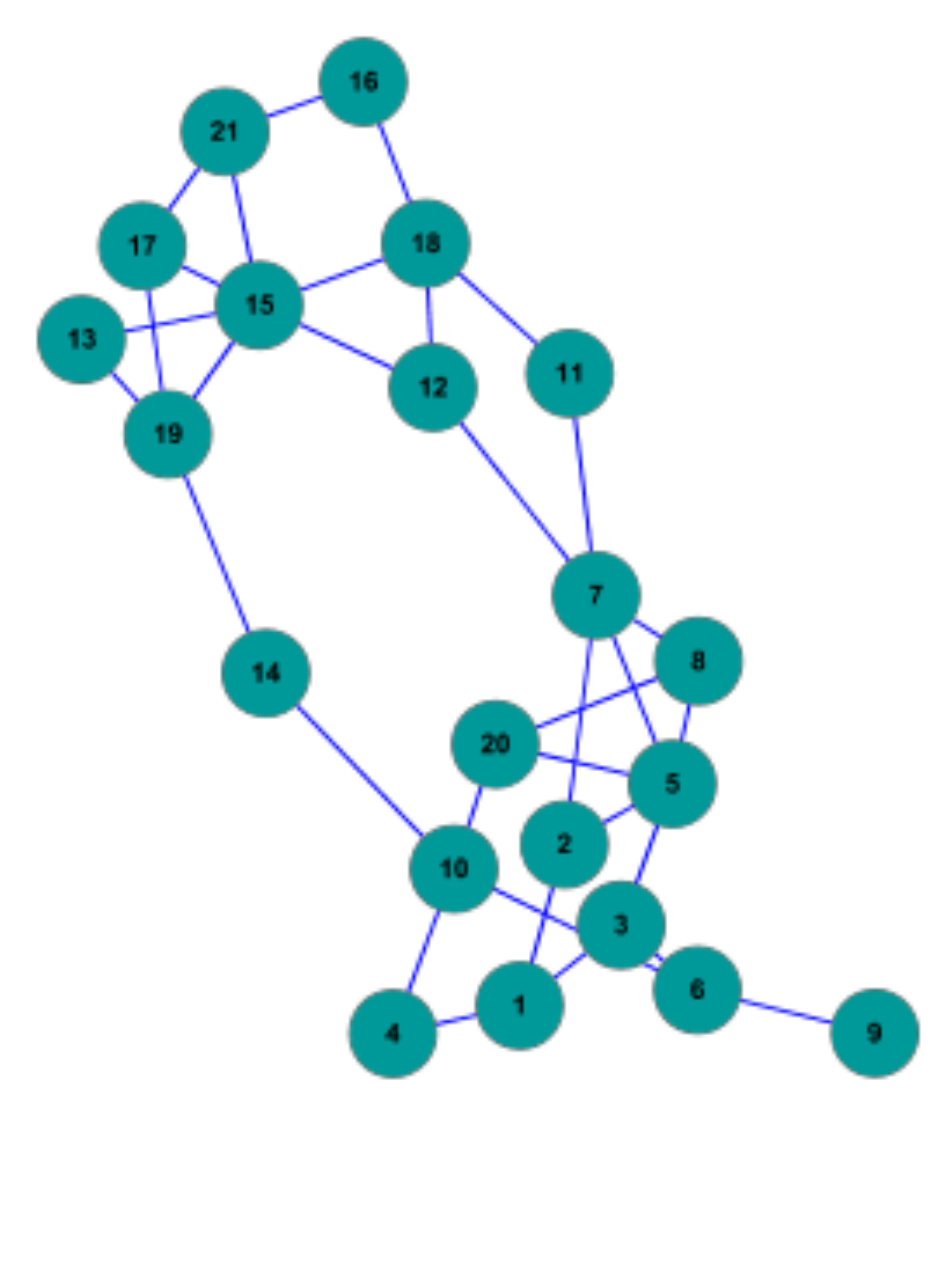}}
  \hspace{0.5cm}
    \subfigure[The six nodes identify by the degree centrality ]{
    \label{Local_networkSS:subfig:b} 
    \centering
    \includegraphics[scale=0.4]{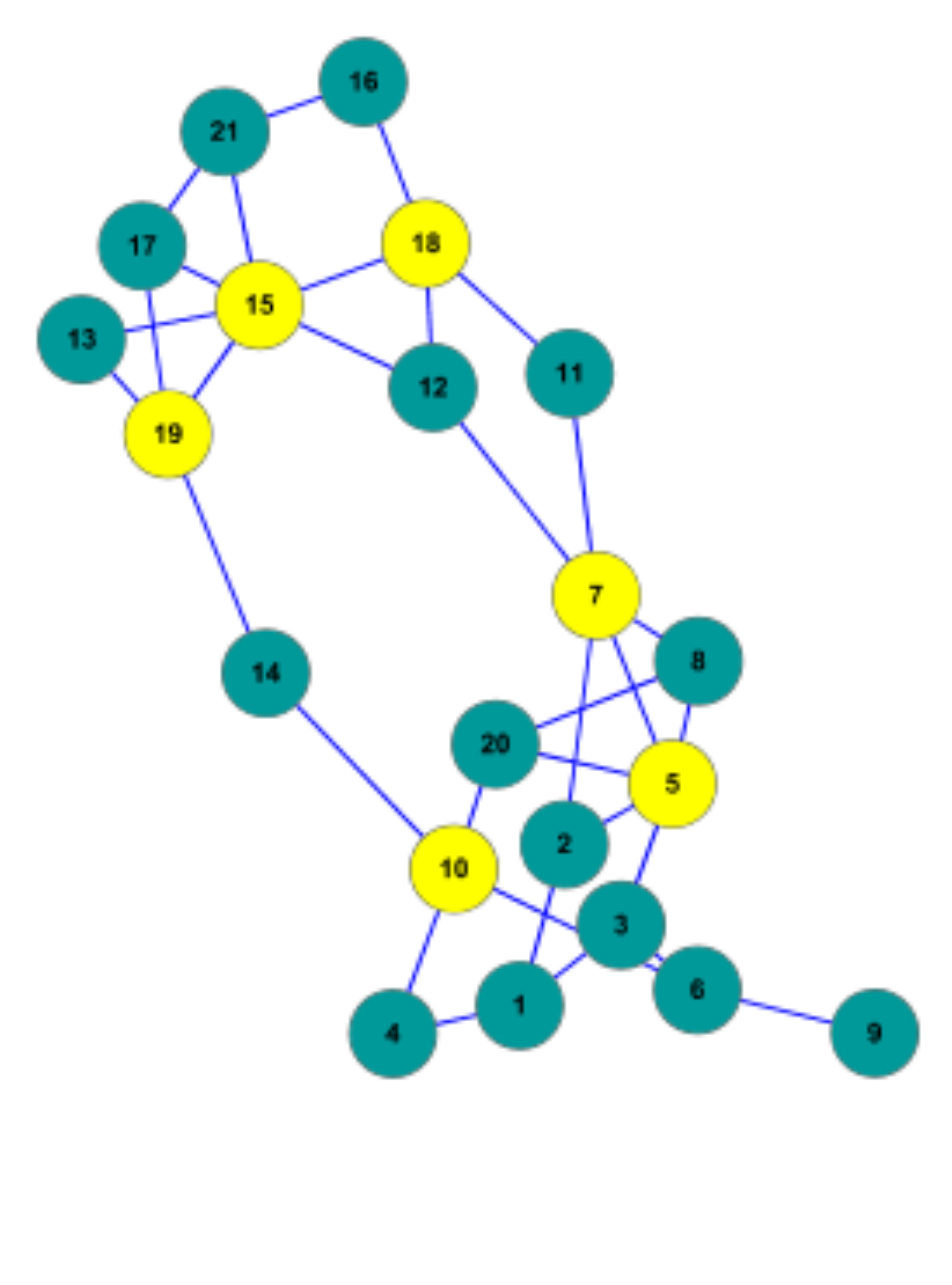}}
  \hspace{0.5cm}
    \subfigure[The six nodes identify by the betweenness centrality]{
    \label{Local_networkSS:subfig:c} 
    \centering
    \includegraphics[scale=0.4]{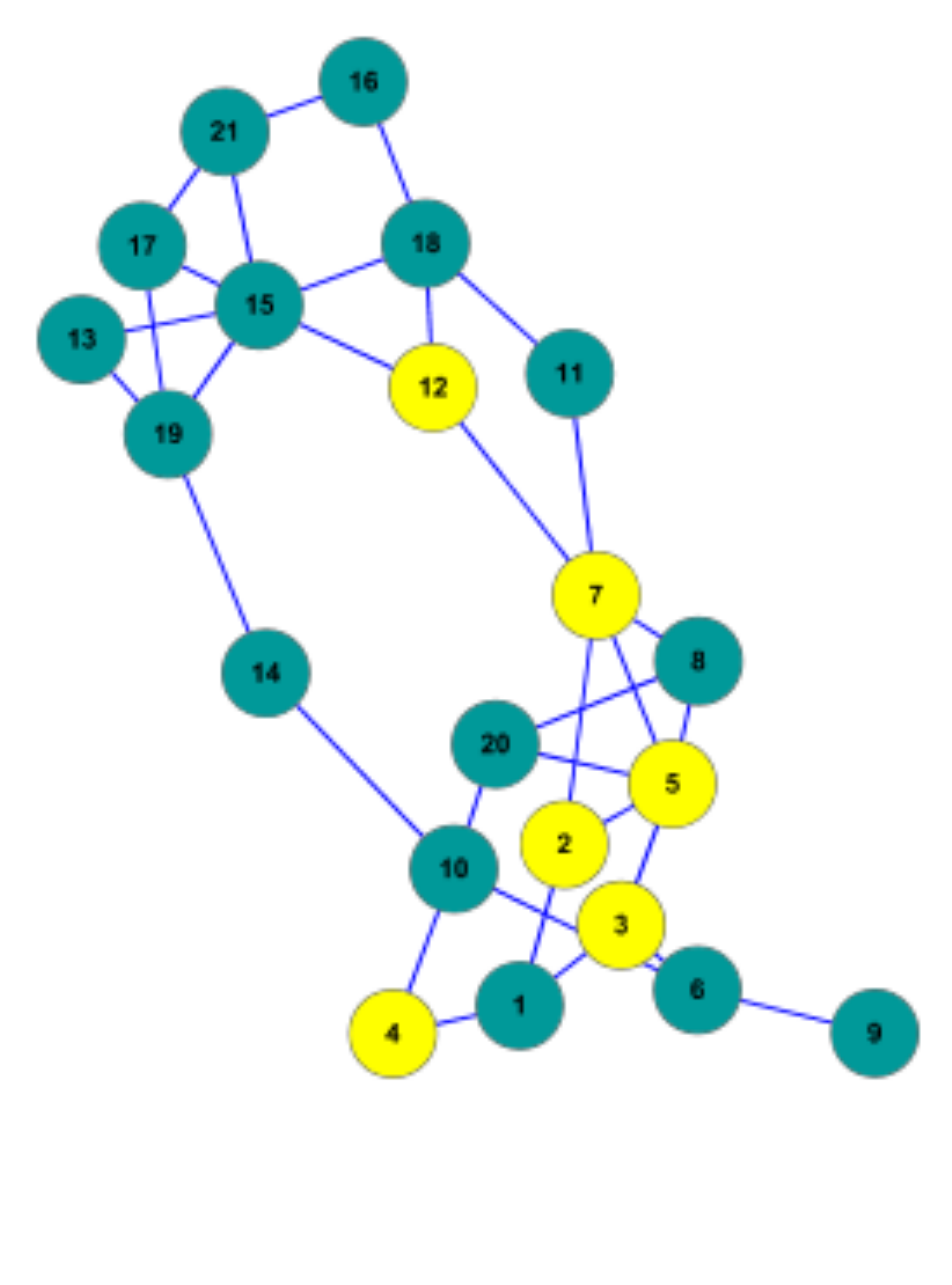}}
  \hspace{0.5cm}
      \subfigure[The six nodes identify by the local structure entropy]{
    \label{Local_networkSS:subfig:d} 
    \centering
    \includegraphics[scale=0.4]{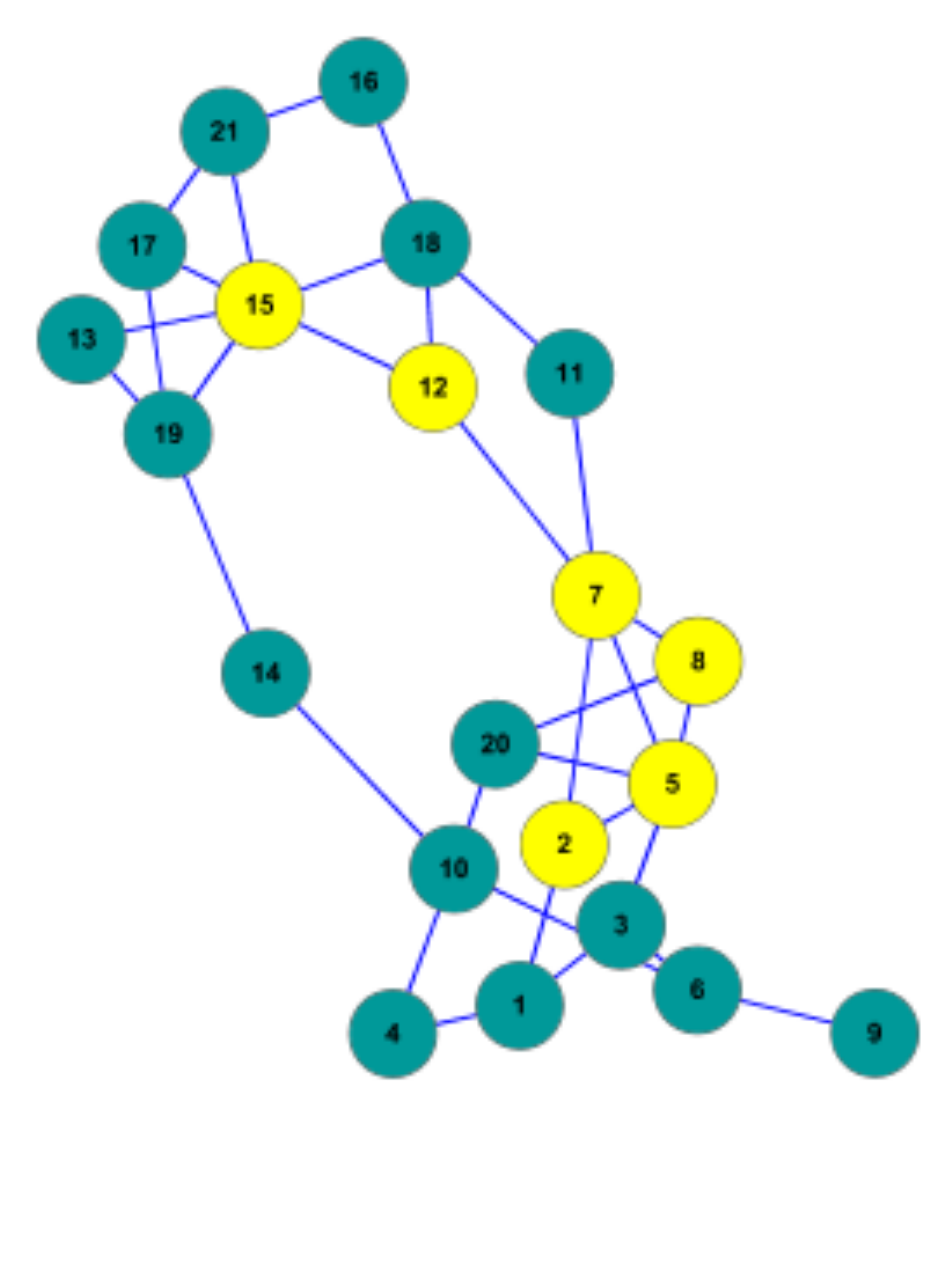}}
  \caption{The different colour of node show the influence of them on the whole network. The node with yellow colour is the six most influential nodes, the nodes with green colour are the normal nodes. }\label{Local_networkSS}
\end{figure}

In the six most influential nodes in the $Network B$ which are identified by the local structure entropy, the top three influential nodes are the same as the nodes identified by the degree centrality and the other three nodes are the same as the nodes identified by the betweenness centrality. It means that, in the $Network B$, the local structure entropy contains some property of the betweenness centrality. In order to show more details of the nodes identifying by the local structure entropy we have point those most influential nodes in the Fig.\ref{Local_networkSS}

From the Fig.\ref{Local_networkSS}, we can see that the nodes identified by the local structure entropy are in the middle of the nodes which have big value of degree. In other words, the nodes identified by the local structure entropy are the nodes which have small value of degree but a series of important neighbour.

\section{Application}
\label{application}
In this section, the local structure entropy of the complex network is used to identify the influence of the nodes in those real networks, such as the Zachary's Karate Club network \cite{uci}, the US-airport network \cite{networkdata}, Email networks \cite{networkdata}, the Germany highway networks \cite{nettt} and the protein-protein interaction network in budding yeast \cite{networkdata}.

\subsection{The property of the local structure entropy}
First, we use the Zachary's Karate Club network \cite{uci} to show the property of the proposed method. The result is shown in the Table\ref{tab:karate_identify}. The top six important nodes have identified by the betweenness centrality, the degree centrality and the local structure entropy.

\begin{table}[htbp]
  \centering
  \caption{The top important nodes in the Zachary's Karate Club network \cite{uci} identify by different methods}
    \begin{tabular}{lcccccc}
    \hline
          & 1     & 2     & 3     & 4     & 5     & 6 \\
    \hline
    Degree  & 34    & 1     & 33    & 3     & 2     & 32 \\
    Betweenness    & 1     & 3     & 34    & 10    & 20    & 32 \\

     ${LE_i}$  & 14    & 9     & 3     & 32    & 31    & 8 \\
    \hline
    \end{tabular}%
  \label{tab:karate_identify}%
\end{table}%

\begin{figure}
    \centering

    \subfigure[The Zachary's Karate Club network \cite{uci} ]{
    \label{karate_E:subfig:a} 
    \centering
    \includegraphics[scale=0.43]{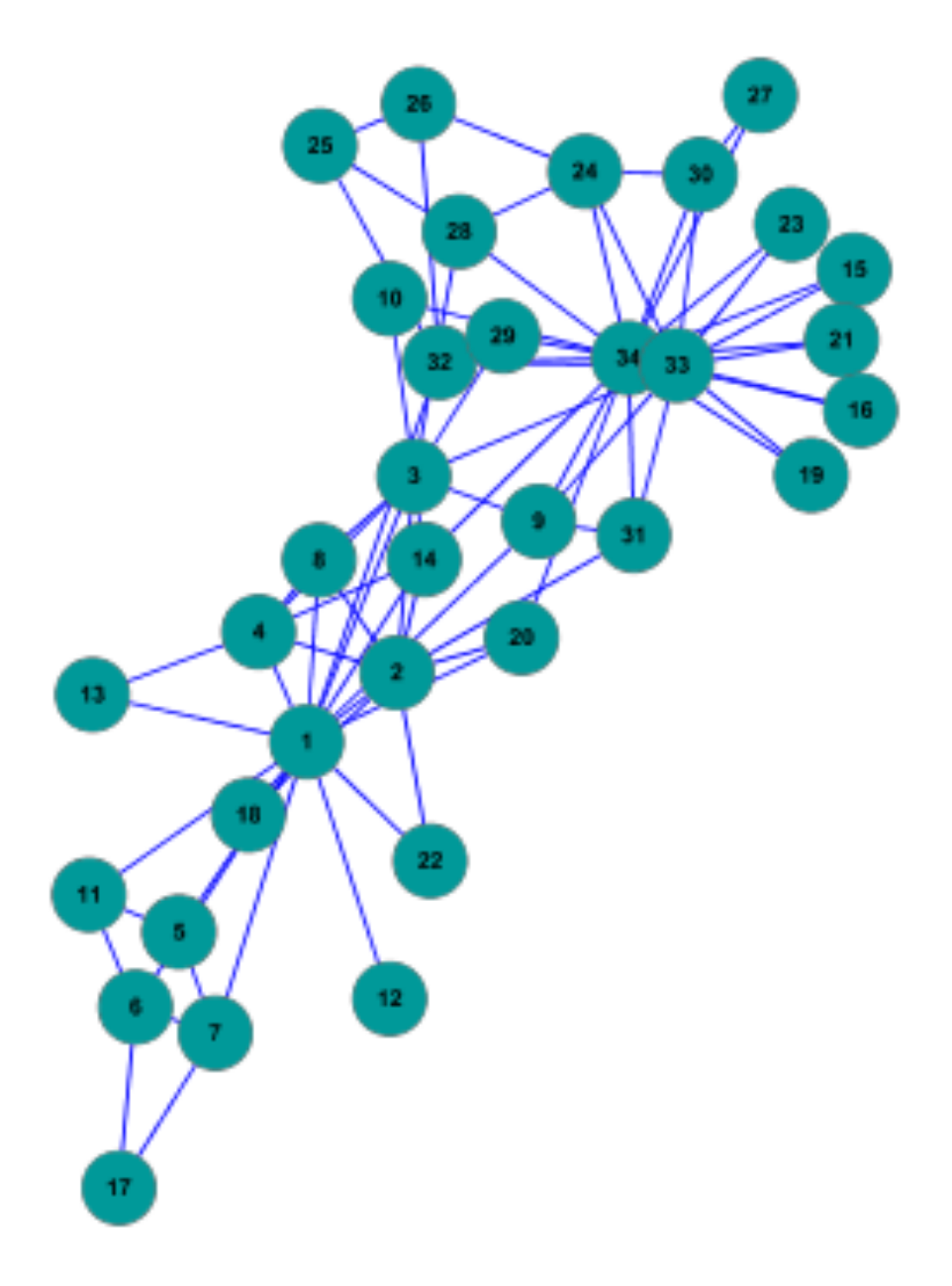}}
  \subfigure[The nodes in the colour of yellow is the top six important nodes in the Zachary's Karate Club network \cite{uci} which are identified by the betweenness centrality ]{
    \label{karate_E:subfig:a} 
    \centering
    \includegraphics[scale=0.4]{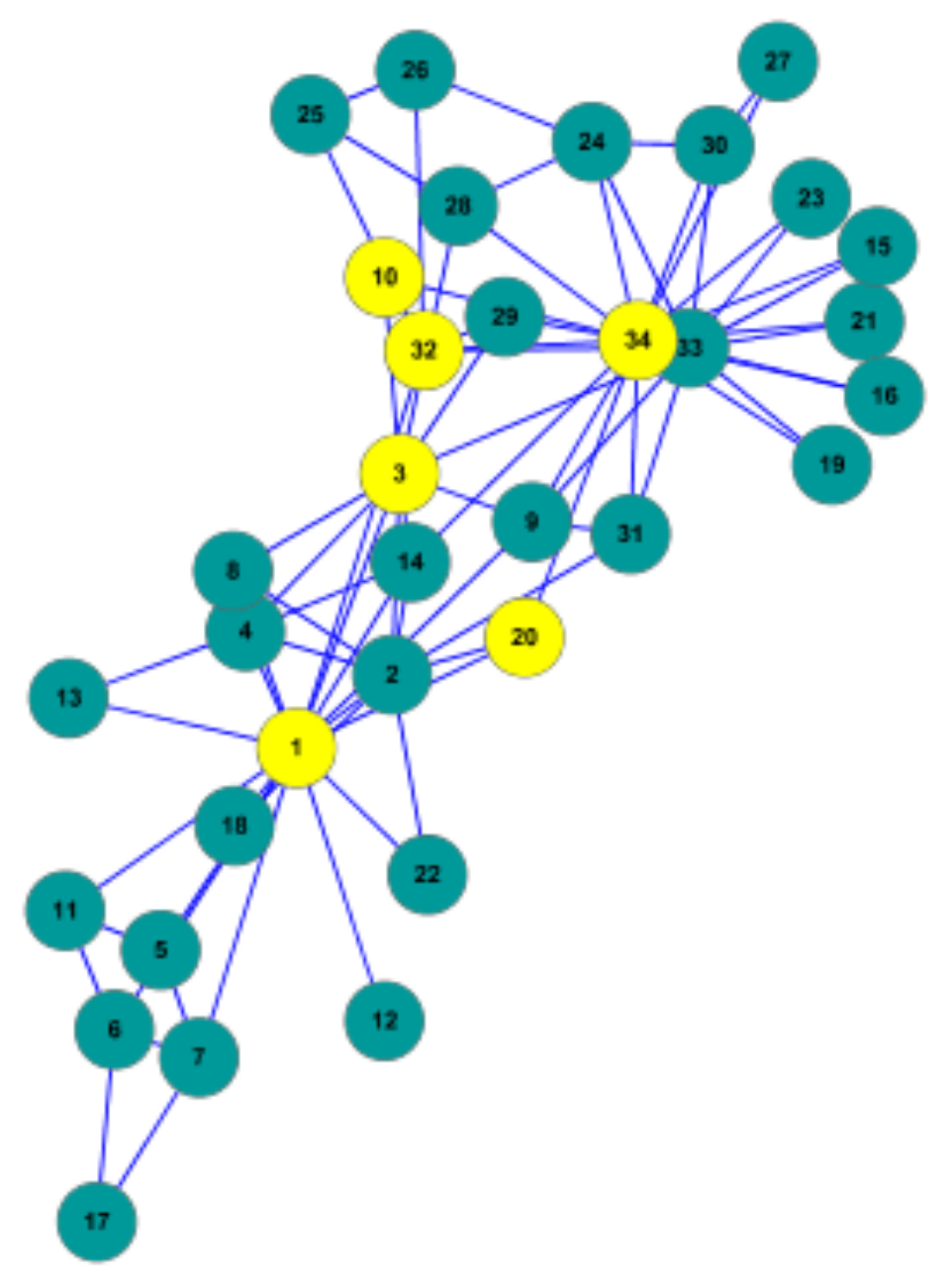}}
    \subfigure[The nodes in the colour of yellow is the top six important nodes in the Zachary's Karate Club network \cite{uci} which are identified by the degree centrality]{
    \label{karate_E:subfig:b} 
    \includegraphics[scale=0.4]{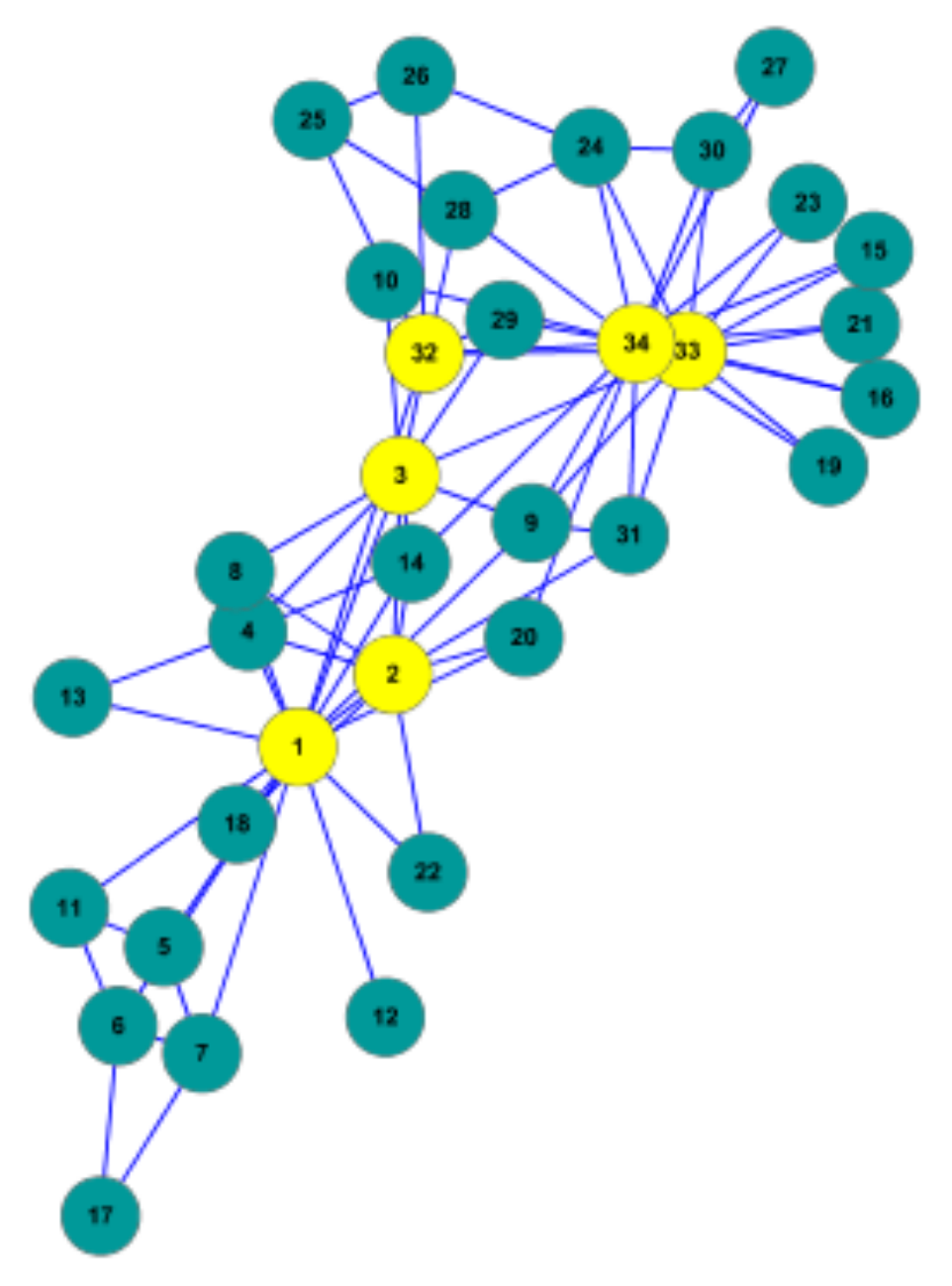}}
    \centering
    \subfigure[The nodes in the colour of yellow is the top six important nodes in the Zachary's Karate Club network \cite{uci} which are identified by the local structure entropy]{
    \label{example:subfig:c} 
    \includegraphics[scale=0.4]{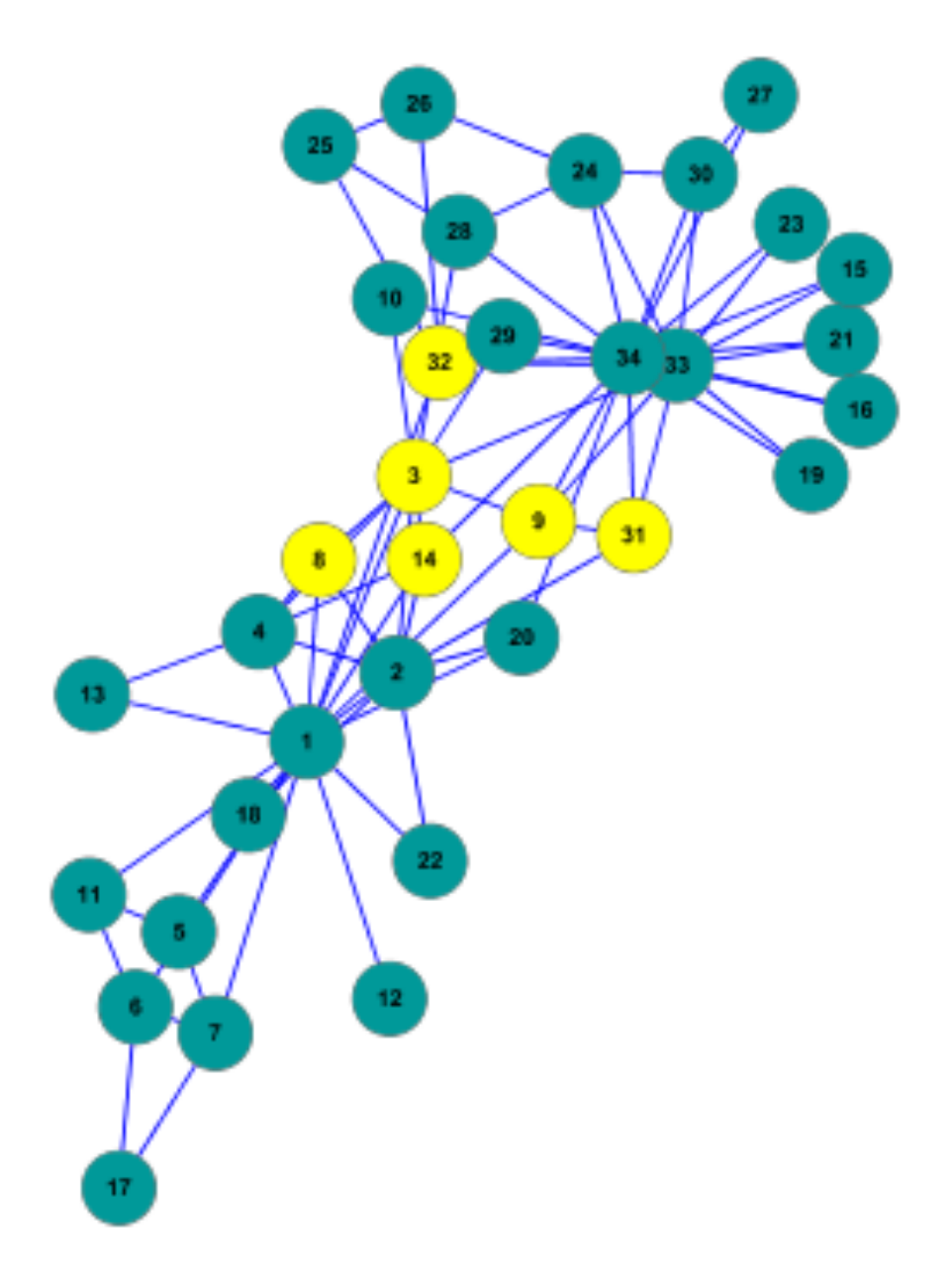}}


  \caption{It is clear that depends on the degree the Zachary's Karate Club network \cite{uci} can be divided into three parts. The degree centrality can identify the central node in the parts. In the figure the local structure entropy can identify the nodes connect the central nodes in the different parts. }\label{karate_E}
\end{figure}

The results of our test on the Zachary's Karate Club network \cite{uci} show that the local structure entropy can identify the nodes connect those node's have big value of degree.

\subsection{The difference between the local structure entropy and the existing methods to identify the influences nodes in the US-airport network \cite{networkdata}}

In order to prove the rationality of the proposed method, we use the SI model to infect the most important nodes identify by different methods in the US-airport network \cite{networkdata}. The results are shown in the Table \ref{tab:different methodl}.

\begin{table}[htbp]
  \centering
  \caption{The most important nodes in the US-airport network \cite{networkdata} identify by different methods}
    \begin{tabular}{llllllllllll}
    \hline
           & 1     & 2     & 3     & 4     & 5     & 6     & 7     & 8     & 9     & 10    & 11   \\
    \hline
    Bet    & 72    & 4     & 151   & 22    & 78    & 103   & 12    & 32    & 116   & 75    & 77   \\
    Degree & 118   & 261   & 255   & 182   & 152   & 230   & 166   & 67    & 112   & 201   & 147 \\
    Local  & 159    & 292    & 172   & 131    & 94    & 109   & 307    & 301    & 310   & 305     & 119   \\
    \hline
    \end{tabular}%
  \label{tab:different methodl}%
\end{table}%

The nodes number in the US-airport network \cite{networkdata} are equal to 332, so each step in the process contains 1 times infection. It means that, each infective node has one chance to infect other neighbour nodes in a step.
The tables show as follows illuminate the proportion of infective nodes in the network in each step. The figures show the process in directly. Because the plot in the step 3 and step 5 is not clear, so we have create two subfigure in every figure.

\begin{table}[htbp]
  \centering
  \caption{The infected proportion of the most important nodes in the US-airport network \cite{networkdata} which are identified by the betweenness centrality}
    \begin{tabular}{lrrrrr}
    \hline
    node  & Step 1 & Step 2 & Step 3 & Step 4& Step 5 \\
    \hline
    72    & 0.005096 & 0.020301 & 0.076009 & 0.185849 & 0.329223 \\
    4     & 0.004500 & 0.008539 & 0.021849 & 0.056163 & 0.114798 \\
    151   & 0.004630 & 0.009069 & 0.025223 & 0.068750 & 0.159895 \\
    22    & 0.003657 & 0.004410 & 0.005539 & 0.008690 & 0.017105 \\
    78    & 0.003955 & 0.011214 & 0.034964 & 0.093919 & 0.180524 \\
    103   & 0.003663 & 0.006925 & 0.019169 & 0.043316 & 0.096943 \\
    12    & 0.003925 & 0.005244 & 0.007030 & 0.010084 & 0.012747 \\
    32    & 0.003581 & 0.004352 & 0.005322 & 0.007355 & 0.012307 \\
    116   & 0.004786 & 0.008009 & 0.017340 & 0.049154 & 0.113822 \\
    75    & 0.005678 & 0.016642 & 0.051352 & 0.135235 & 0.241247 \\
    77    & 0.004304 & 0.010663 & 0.034877 & 0.089289 & 0.168337 \\
    \hline
    \end{tabular}%
  \label{tab:us-air-betweenness}%
\end{table}%

\begin{figure}
    \centering

    \subfigure[The process of the infection the nodes in the US-airport network \cite{networkdata} which are identified by the betweenness centrality ]{
    \label{Usair-infect-bet1} 
    \centering
    \includegraphics[scale=0.585]{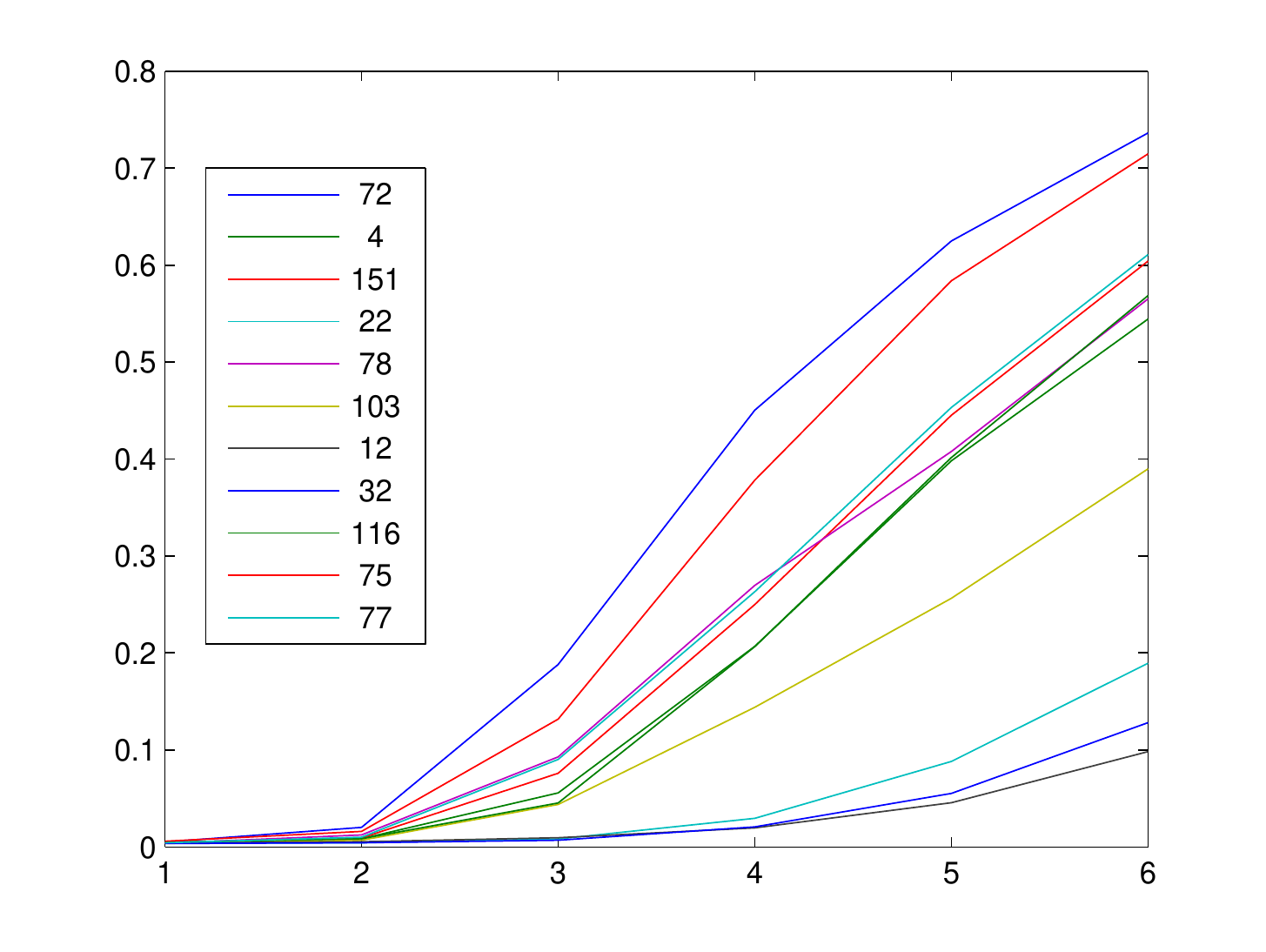}}
  \subfigure[The details in the step 3. ]{
    \label{Usair-infect-bet:a} 
    \centering
    \includegraphics[scale=0.28]{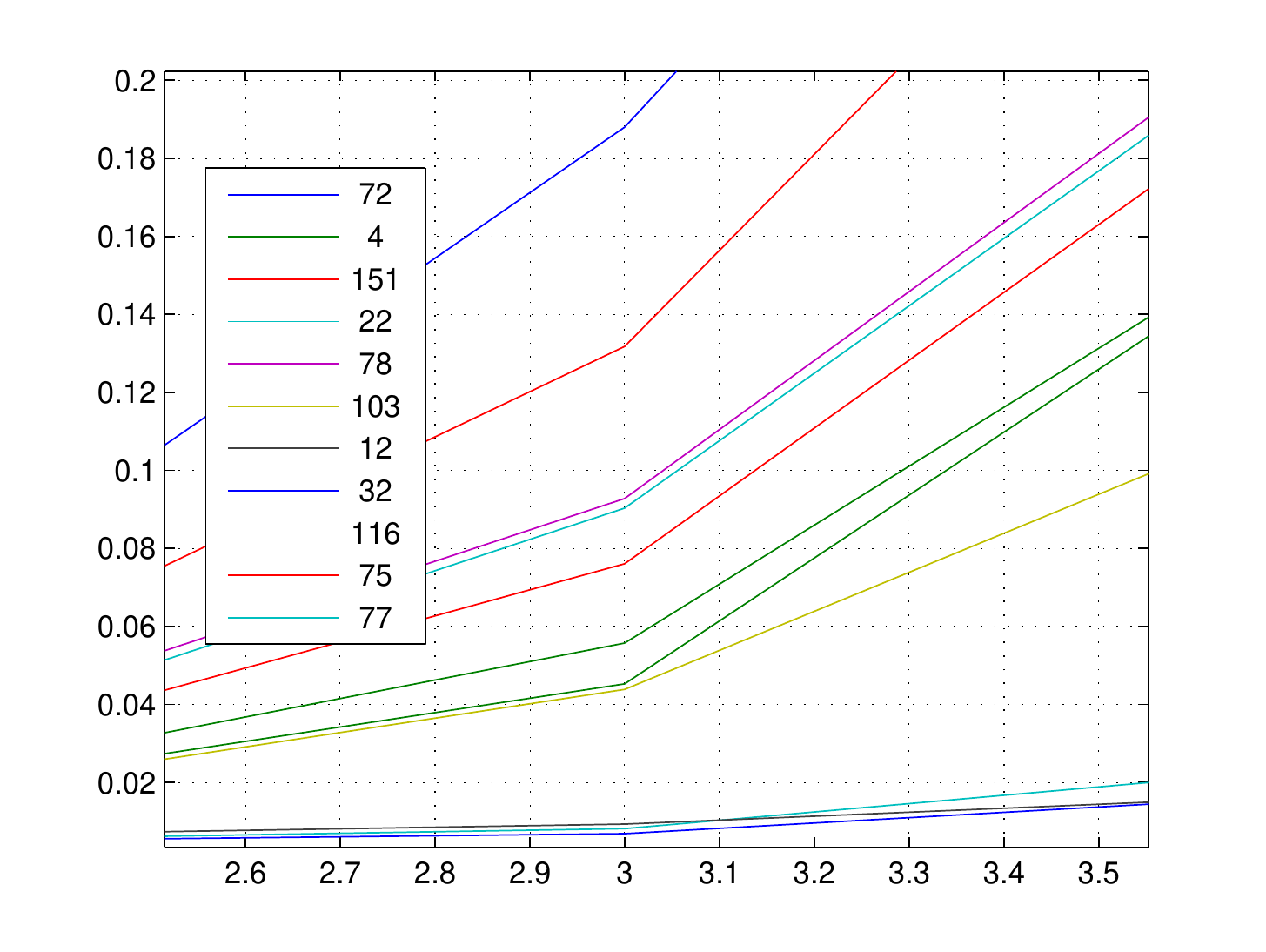}}
    \subfigure[The details in the step 5.]{
    \label{Usair-infect-bet:b} 
    \includegraphics[scale=0.28]{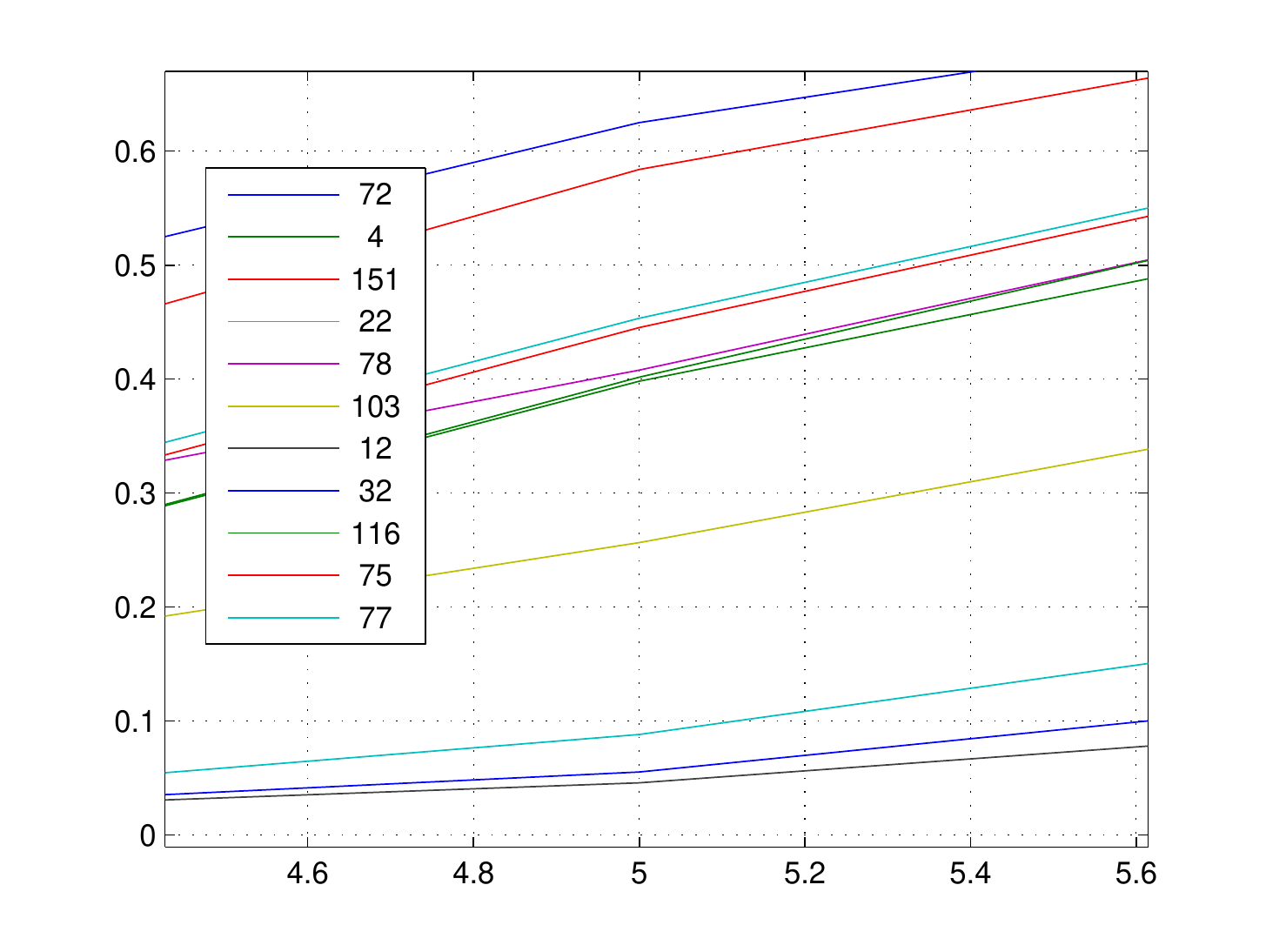}}
    \centering
  \caption{The infected proportion of the most important nodes in the US-airport network \cite{networkdata} which are identified by the betweenness centrality. }\label{Usair-infect-bet}
\end{figure}

The Fig.\ref{Usair-infect-bet} and Table \ref{tab:us-air-betweenness} show the proportion of the infective nodes in the network of each step. The infection source nodes show in the figure and table are identify by the betweenneess centrality. The results show that in the first step, each node which has been identified by the betweenness centrality has a small percentage of infection. It means that most of the nodes have a small value of degree, so that they can not infect many nodes in the first step. In the step 2, step 3, step 4 and step 5, most of the nodes which are identified by the betweenness centrality has a small percentage of infection too. It means that depends on the SI model, the nodes seems not so important to influence the whole network.

\begin{table}[htbp]
  \centering
  \caption{The infected proportion of the most important nodes in the US-airport network \cite{networkdata} which are identified by the degree centrality}
    \begin{tabular}{lrrrrr}
    \hline
    node  & Step 1 & Step 2 & Step 3 & Step 4& Step 5 \\
    \hline
    118   & 0.044732 & 0.153503 & 0.336256 & 0.487346 & 0.588919 \\
    \hline
    261   & 0.038596 & 0.138346 & 0.316979 & 0.473256 & 0.578030 \\
    255   & 0.033506 & 0.126413 & 0.307741 & 0.462443 & 0.566184 \\
    182   & 0.031145 & 0.122277 & 0.302256 & 0.460419 & 0.570030 \\
    152   & 0.031449 & 0.118720 & 0.295136 & 0.450786 & 0.560108 \\
    230   & 0.029154 & 0.112485 & 0.287473 & 0.443961 & 0.555151 \\
    166   & 0.028648 & 0.115937 & 0.294364 & 0.454518 & 0.568265 \\
    67    & 0.026265 & 0.111458 & 0.293892 & 0.457370 & 0.571027 \\
    112   & 0.024096 & 0.106078 & 0.280491 & 0.444943 & 0.557774 \\
    201   & 0.023587 & 0.100617 & 0.272750 & 0.444717 & 0.562714 \\
    147   & 0.023133 & 0.099148 & 0.269545 & 0.428602 & 0.544455 \\
    \hline
    \end{tabular}%
  \label{tab:us-air-degree}%
\end{table}%

\begin{figure}
    \centering

    \subfigure[The process of the infection the nodes in the US-airport network \cite{networkdata} which are identified by the degree centrality ]{
    \label{Usair-infect-degree1} 
    \centering
    \includegraphics[scale=0.585]{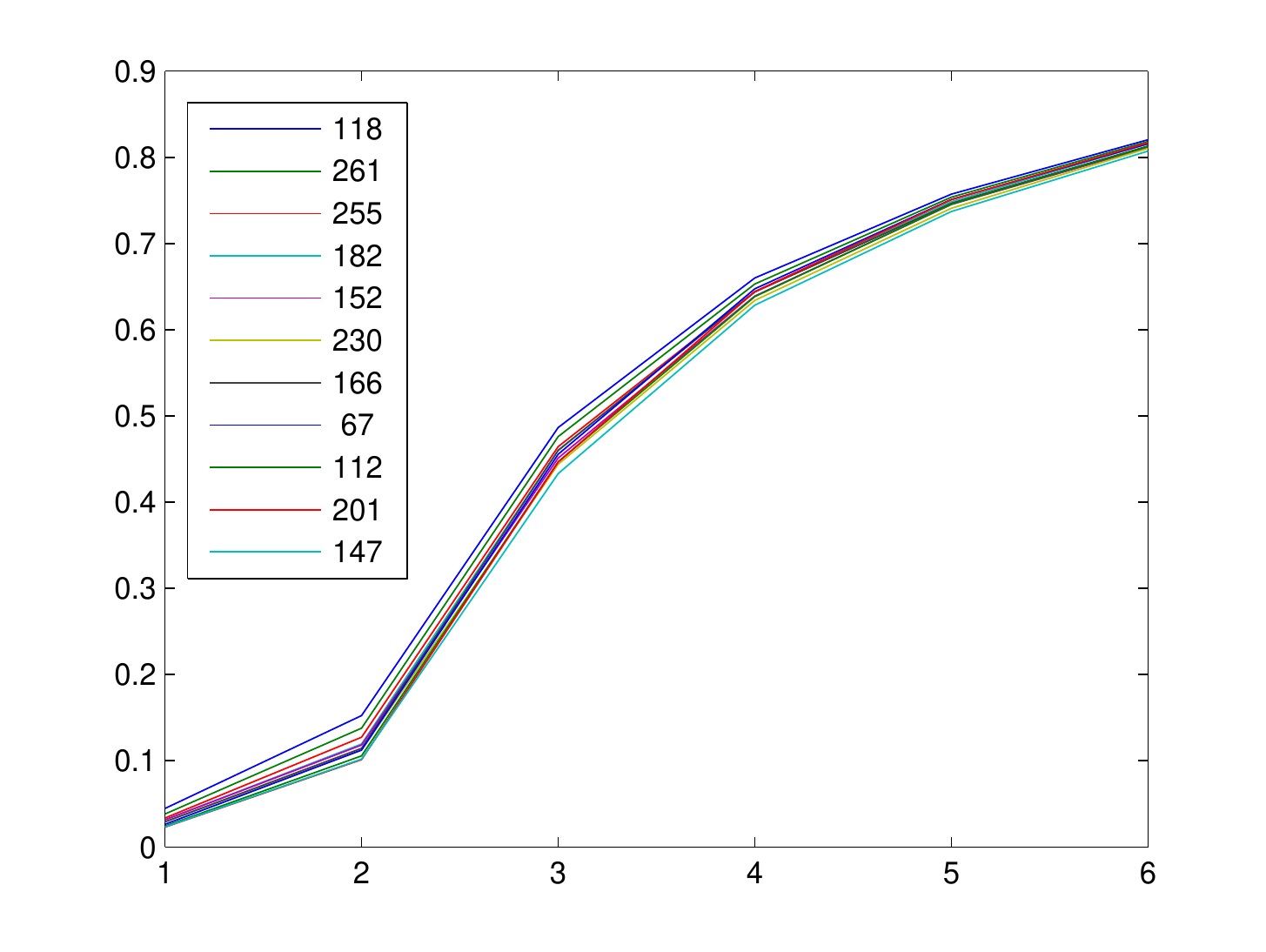}}
  \subfigure[The details in the step 3. ]{
    \label{Usair-infect-degree:a} 
    \centering
    \includegraphics[scale=0.28]{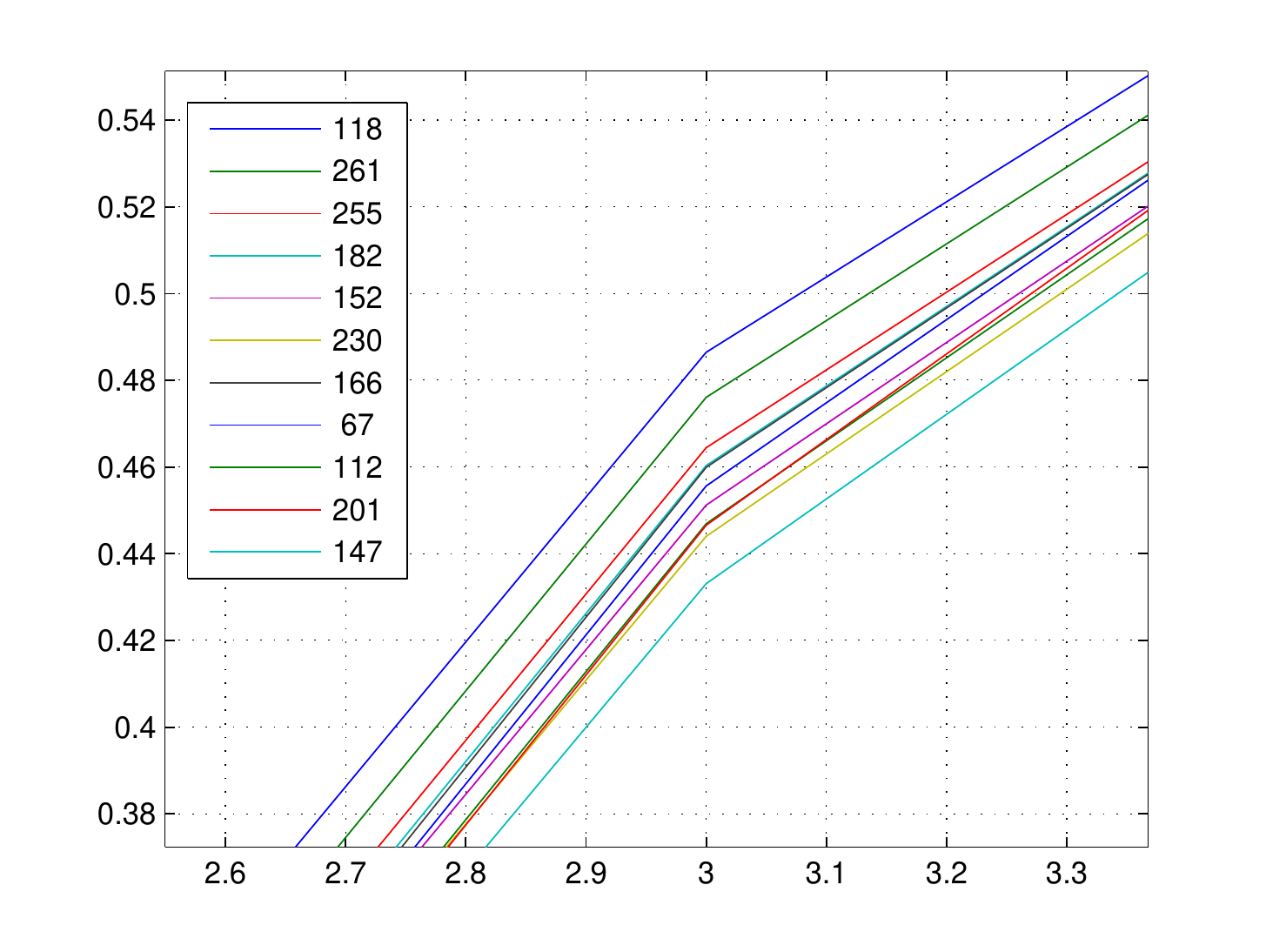}}
    \subfigure[The details in the step 5.]{
    \label{Usair-infect-degree:b} 
    \includegraphics[scale=0.28]{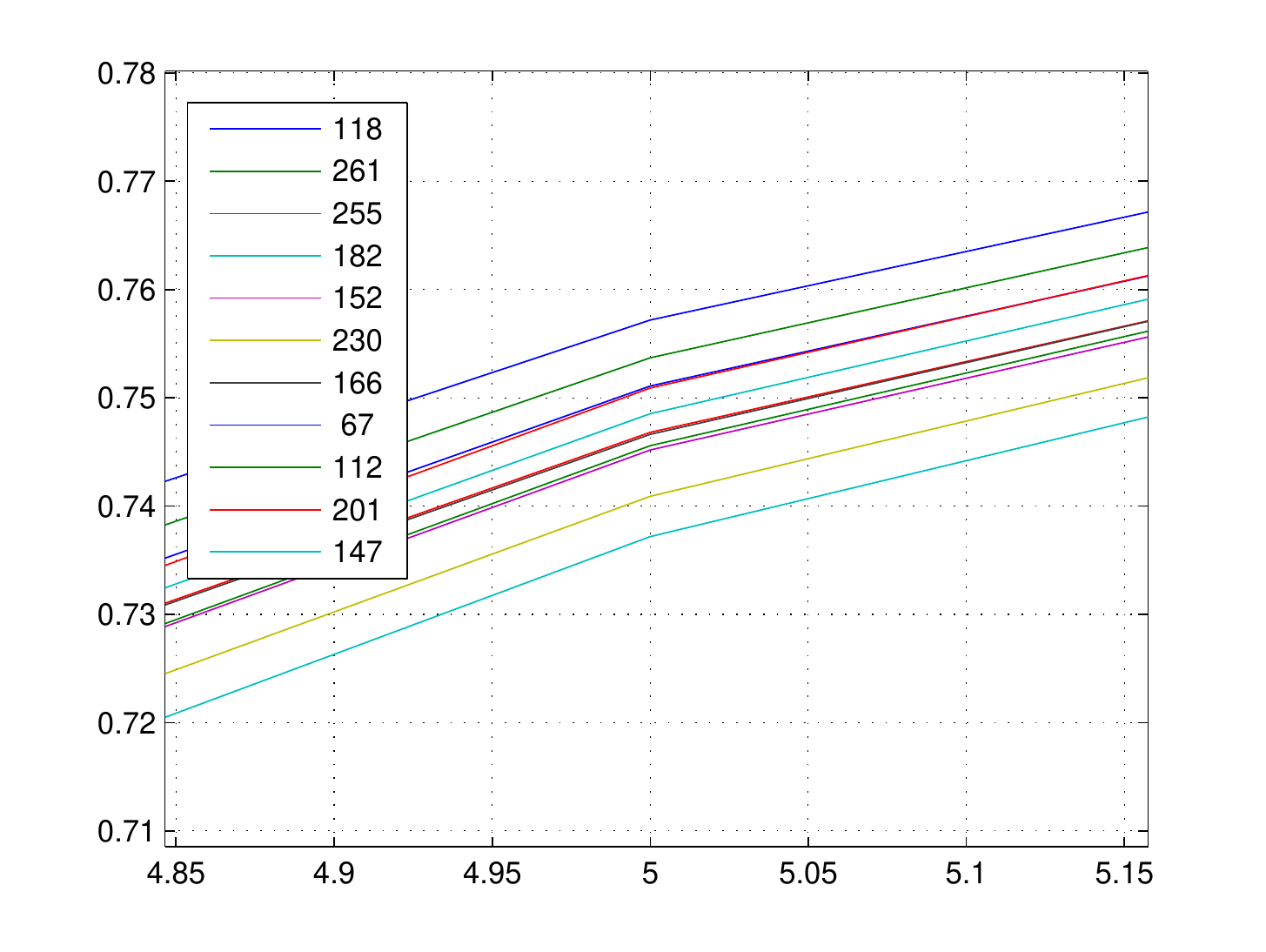}}
    \centering
  \caption{The infected proportion of the most important nodes in the US-airport network \cite{networkdata} which are identified by the degree centrality. }\label{Usair-infect-degree}
\end{figure}
In the Fig.\ref{Usair-infect-degree} and the Table \ref{tab:us-air-degree}, the infection source nodes are identified by the degree centrality. In the first step, each node has a large percentage of infection. It means that the nodes which are identified by the degree centrality have a large value of degree and most of them can infect a lot of nodes in the network.

\begin{table}[htbp]
  \centering
  \caption{The infected proportion of the most important nodes in the US-airport network \cite{networkdata} which are identified by the local structure entropy}
    \begin{tabular}{lrrrrr}
    \hline
    node  & Step 1 & Step 2 & Step 3 & Step 4& Step 5 \\
    \hline
    159   & 0.011443 & 0.065630 & 0.208238 & 0.381780 & 0.515051 \\
    292   & 0.013139 & 0.072611 & 0.221934 & 0.395461 & 0.522292 \\
    172   & 0.013262 & 0.073033 & 0.228202 & 0.402955 & 0.526536 \\
    131   & 0.016518 & 0.084804 & 0.247870 & 0.417898 & 0.536590 \\
    94    & 0.013295 & 0.072367 & 0.217744 & 0.400922 & 0.524855 \\
    109   & 0.017964 & 0.089988 & 0.255310 & 0.424078 & 0.541928 \\
    307   & 0.011566 & 0.063181 & 0.203738 & 0.371910 & 0.501786 \\
    301   & 0.015482 & 0.078120 & 0.232105 & 0.405461 & 0.526636 \\
    310   & 0.012831 & 0.069651 & 0.211166 & 0.387636 & 0.516117 \\
    305   & 0.008657 & 0.049741 & 0.170093 & 0.333994 & 0.475587 \\
    119   & 0.011828 & 0.064617 & 0.199910 & 0.368413 & 0.502720 \\
    \hline
    \end{tabular}%
  \label{tab:us-air-local}%
\end{table}%

\begin{figure}
    \centering

    \subfigure[The process of the infection the nodes in the US-airport network \cite{networkdata} which are identified by the local structure entropy ]{
    \label{Usair-infect-local1} 
    \centering
    \includegraphics[scale=0.585]{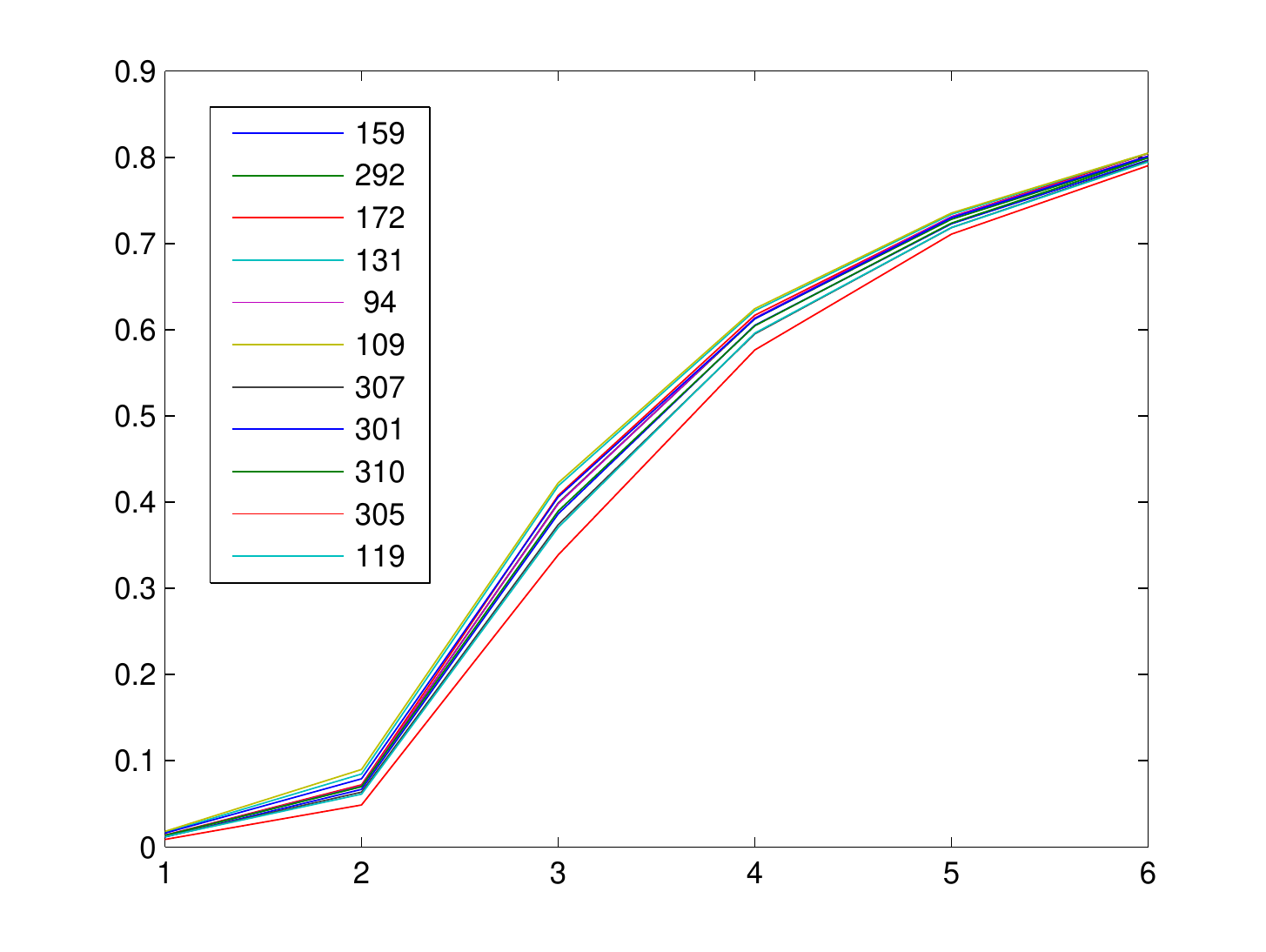}}
  \subfigure[The details in the step 3. ]{
    \label{Usair-infect-local:a} 
    \centering
    \includegraphics[scale=0.28]{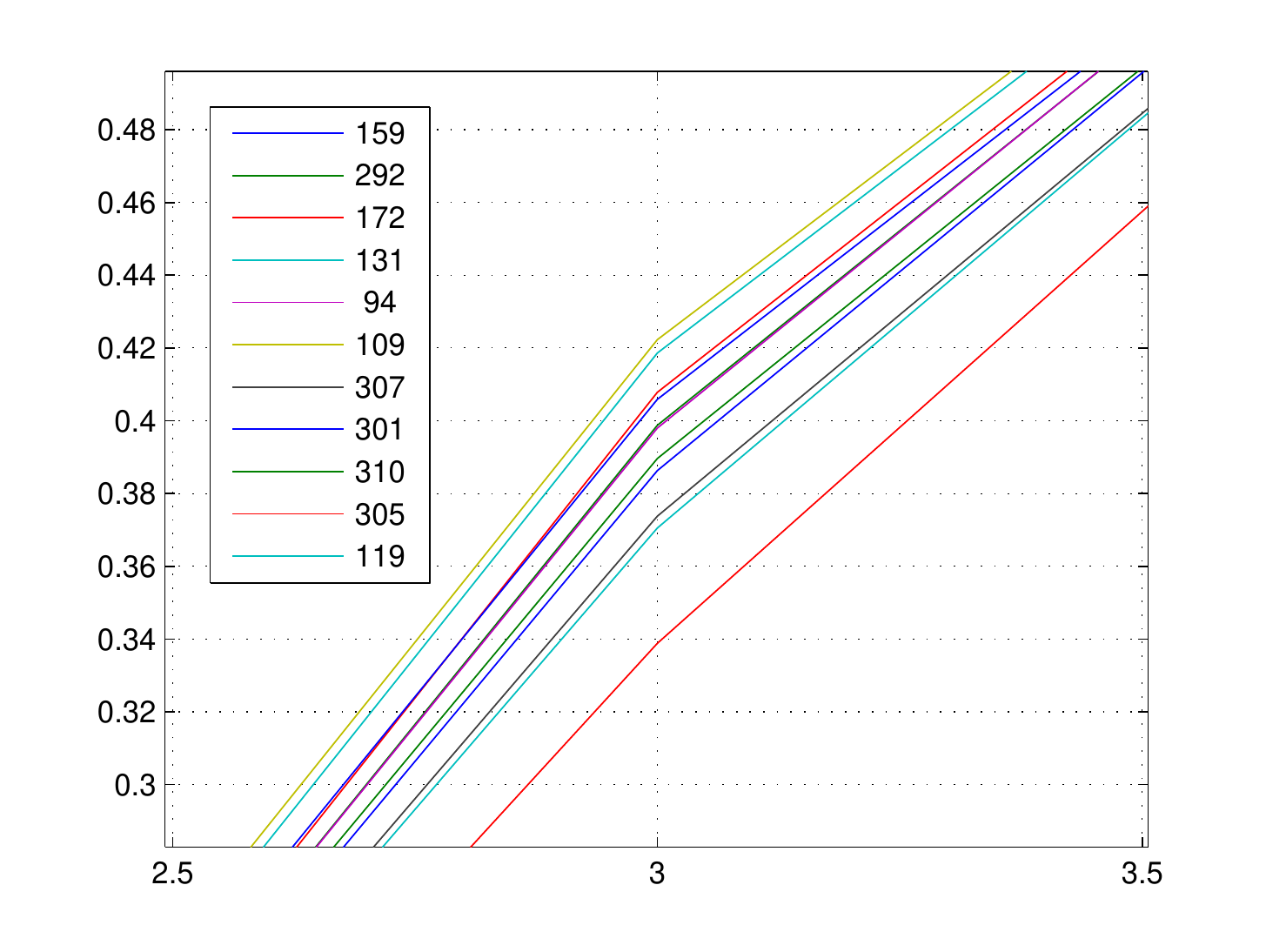}}
    \subfigure[The details in the step 5.]{
    \label{Usair-infect-local:b} 
    \includegraphics[scale=0.28]{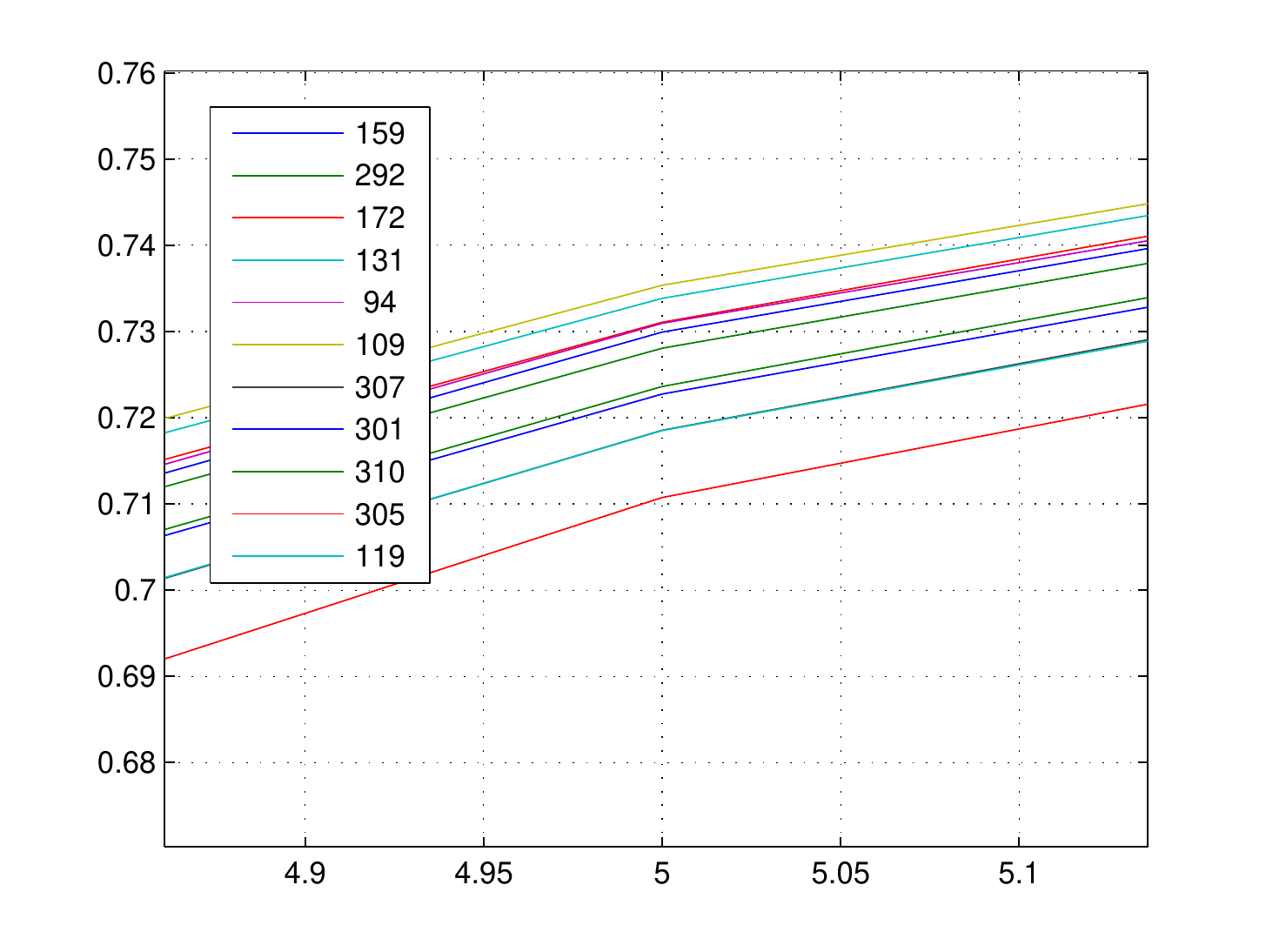}}
    \centering
  \caption{The infected proportion of the most important nodes in the US-airport network \cite{networkdata} which are identified by the local structure entropy. }\label{Usair-infect-local}
\end{figure}
In the Fig.\ref{Usair-infect-local} and the Table \ref{tab:us-air-local}, the infection source nodes are identified by the local structure entropy. In the first step, most of them have infected a small percentage of the nodes in the networks. It means all of them have a small value of degree. However, follow the continue of the process of infection, the percentage of the infective nodes in the network is growing and most of the nodes in the network can be infected. It means that, depends on the SI model, those nodes have an larger influence in the US-airport network \cite{networkdata}.

\subsection{The influential nodes in the reals networks}

In this subsection the influential nodes in the Email networks \cite{networkdata}, the Germany highway networks \cite{nettt} and the protein-protein interaction network in budding yeast \cite{networkdata} are identified by the local structure entropy. The results are shown as follows.

\begin{table}[htbp]
  \centering
  \caption{The top 11 influential nodes in the real networks which are identified by the local structure entropy}
    \begin{tabular}{lccccccccccc}
    \hline
    Node  & 1     & 2     & 3     & 4     & 5     & 6     & 7     & 8     & 9     & 10    & 11 \\
    \hline
    Email \cite{networkdata} & 105   & 3     & 39    & 16    & 42    & 54    & 210   & 390   & 50    & 332   & 9 \\
    Yeast  \cite{networkdata}  & 944   & 942   & 941   & 939   & 935   & 928   & 185   & 184   & 940   & 938   & 927 \\
    High  \cite{nettt} & 219   & 393   & 698   & 217   & 404   & 450   & 543   & 267   & 331   & 763   & 198 \\
    \hline
    \end{tabular}%
  \label{tab:Reals_networks_results}%
\end{table}%

\begin{table}[htbp]
  \centering
  \caption{The infected proportion of the most important nodes in the Email networks \cite{networkdata} which are identified by the local structure entropy}
    \begin{tabular}{lrrrrrr}
    \hline
    node  & Step 1 & Step 5 & Step 10 & Step 15 & Step 20 & Step 25 \\
    \hline
    105   & 0.007169 & 0.024470 & 0.142155 & 0.405688 & 0.642948 & 0.777696 \\
    3     & 0.004334 & 0.014746 & 0.105977 & 0.354055 & 0.613838 & 0.765627 \\
    39    & 0.003427 & 0.011937 & 0.090041 & 0.319729 & 0.589242 & 0.753282 \\
    16    & 0.005359 & 0.017822 & 0.107098 & 0.326679 & 0.583421 & 0.7471 \\
    42    & 0.005358 & 0.018813 & 0.121579 & 0.381916 & 0.629185 & 0.772535 \\
    54    & 0.003673 & 0.013206 & 0.092159 & 0.330798 & 0.598809 & 0.757853 \\
    210   & 0.003037 & 0.010274 & 0.076484 & 0.280225 & 0.550041 & 0.725693 \\
    390   & 0.003105 & 0.010645 & 0.075457 & 0.277412 & 0.534052 & 0.717746 \\
    50    & 0.003186 & 0.011180 & 0.083778 & 0.314853 & 0.582951 & 0.747732 \\
    332   & 0.004306 & 0.014396 & 0.096790 & 0.325471 & 0.587406 & 0.75068 \\
    9     & 0.002402 & 0.008184 & 0.063924 & 0.247455 & 0.521470 & 0.707512 \\
    \hline
    \end{tabular}%
  \label{tab:Email-local}%
\end{table}%

\begin{figure}
    \centering

    \subfigure[The process of the infection the nodes in the Email networks \cite{networkdata} which are identified by the local structure entropy]{
    \label{Email-infect-local1} 
    \centering
    \includegraphics[scale=0.585]{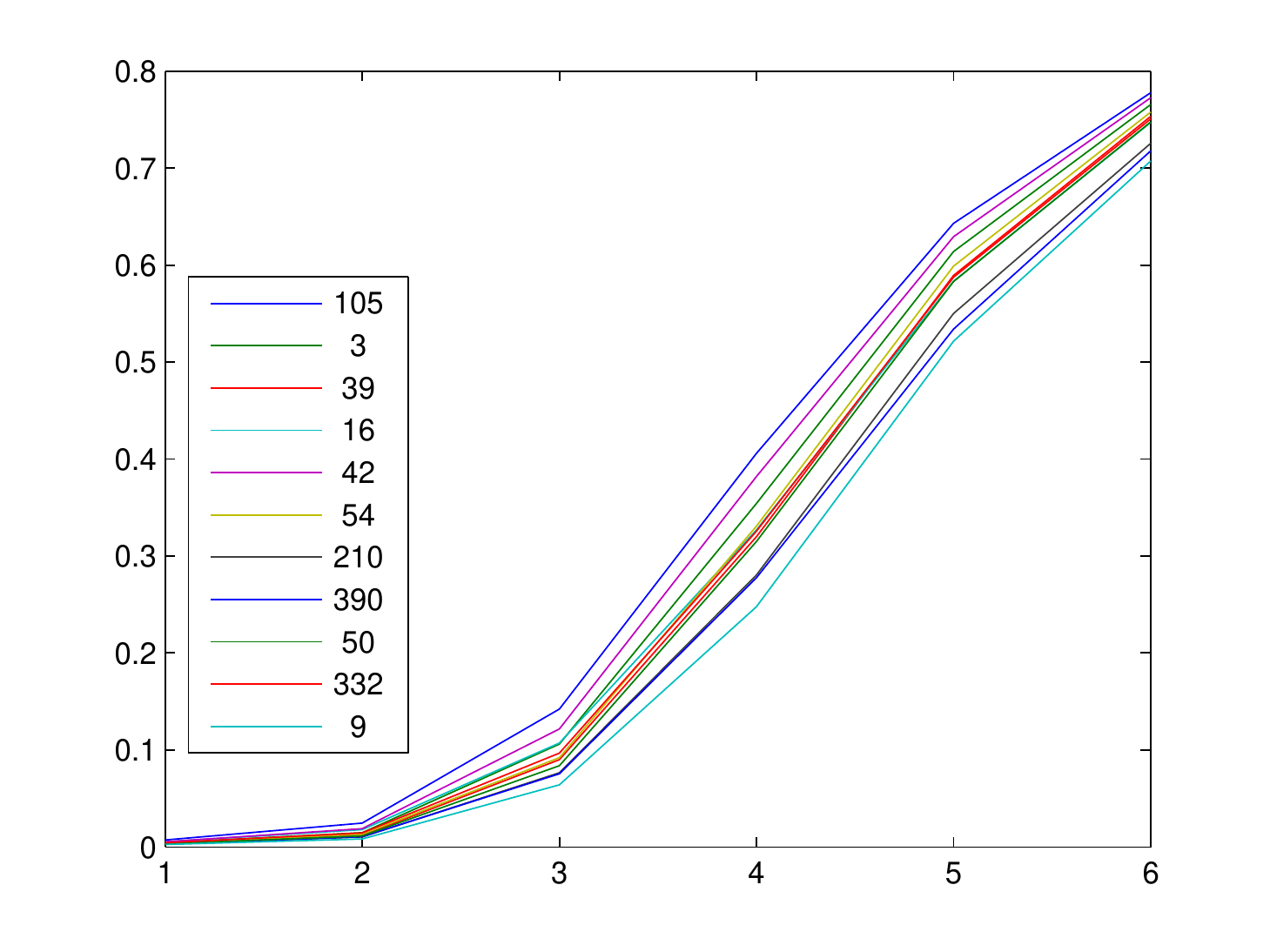}}
  \subfigure[The details in the step 3. ]{
    \label{Email-infect-local:a} 
    \centering
    \includegraphics[scale=0.28]{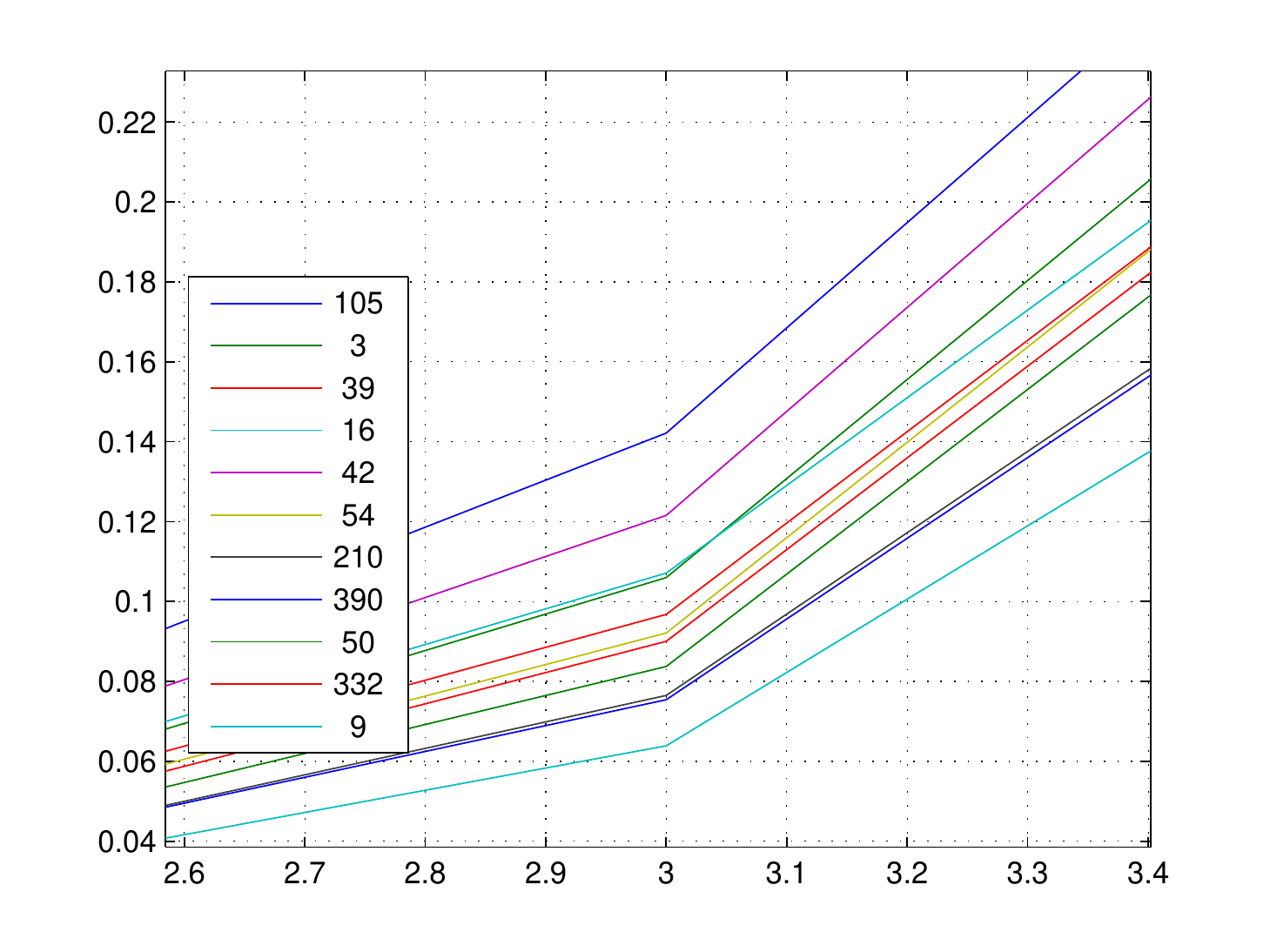}}
    \subfigure[The details in the step 5.]{
    \label{Email-infect-local:b} 
    \includegraphics[scale=0.28]{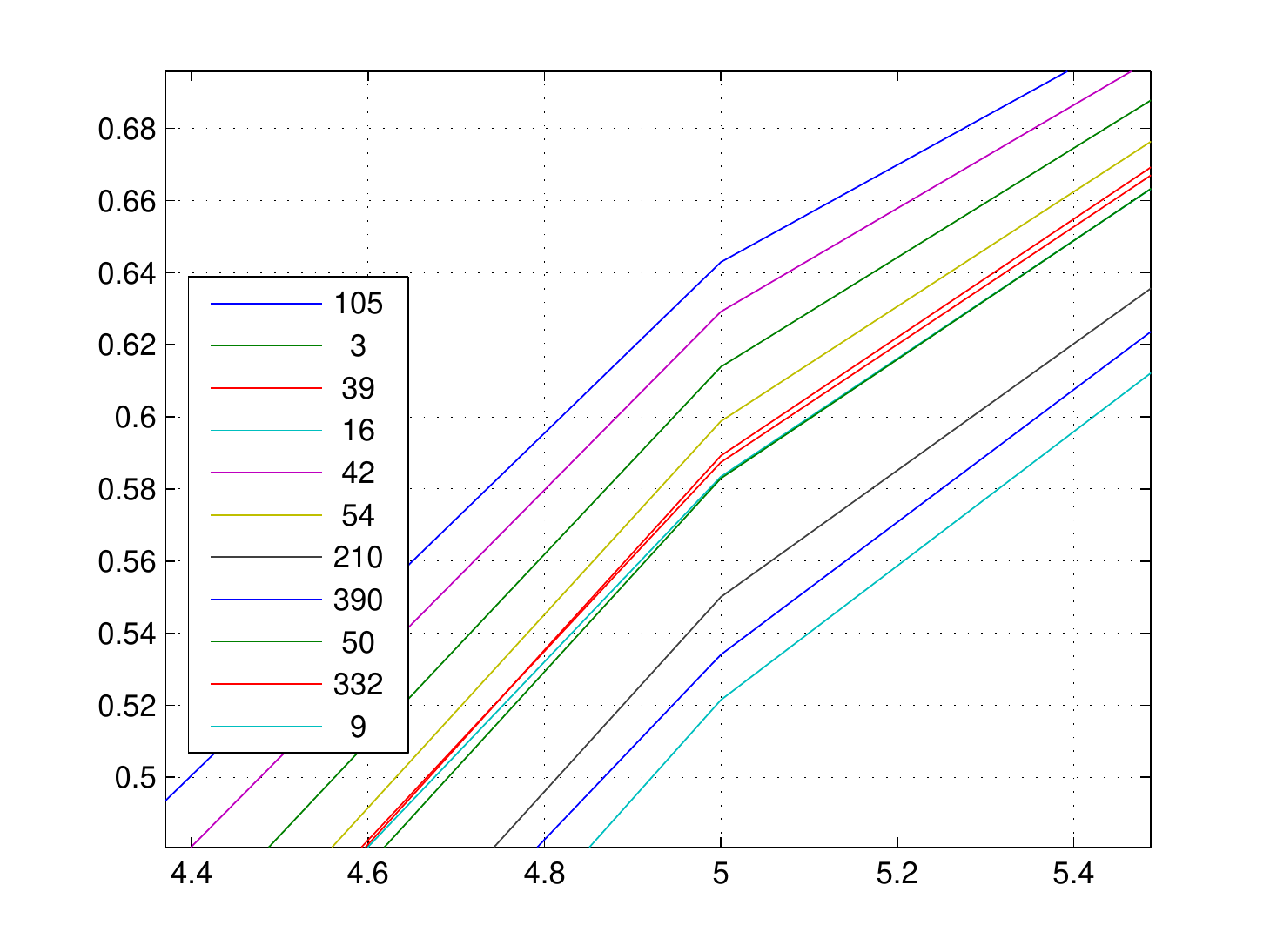}}
    \centering
  \caption{The infected proportion of the most important nodes in the Email networks \cite{networkdata} which are identified by the local structure entropy. }\label{Email-infect-local}
\end{figure}

The process of the infection in the Email networks \cite{networkdata} is shown in the Table \ref{tab:Email-local} and the Fig.\ref{Email-infect-local}. The infection source nodes in the Email networks \cite{networkdata} are identified by the local structure entropy. The results show that each influential node in the Email networks \cite{networkdata} which is identified by the local structure entropy can infect most of the nodes after 25 times infection.

\begin{table}[htbp]
  \centering
  \caption{The infected proportion of the most important nodes in the Germany highway networks \cite{nettt} which are identified by the local structure entropy}
    \begin{tabular}{lrrrrrr}
    \hline
    node  & Step 1 & Step 20 & Step 40 & Step 60 & Step 80 & Step 100 \\
    \hline
    219   & 0.001472 & 0.040973 & 0.132066 & 0.266482 & 0.413909 & 0.548330 \\
    393   & 0.001357 & 0.027193 & 0.071162 & 0.149539 & 0.265228 & 0.414303 \\
    698   & 0.001394 & 0.034184 & 0.128745 & 0.266596 & 0.421396 & 0.560382 \\
    217   & 0.001533 & 0.032535 & 0.107636 & 0.228053 & 0.372823 & 0.515809 \\
    404   & 0.001166 & 0.023276 & 0.073836 & 0.164771 & 0.295297 & 0.441245 \\
    450   & 0.001193 & 0.030004 & 0.116621 & 0.251347 & 0.401224 & 0.540087 \\
    543   & 0.001378 & 0.028094 & 0.098118 & 0.204311 & 0.354270 & 0.501434 \\
    267   & 0.001356 & 0.033911 & 0.118902 & 0.253054 & 0.397814 & 0.535933 \\
    331   & 0.001461 & 0.030902 & 0.119276 & 0.253339 & 0.399676 & 0.540528 \\
    763   & 0.001253 & 0.026887 & 0.106028 & 0.241551 & 0.401823 & 0.542743 \\
    198   & 0.001217 & 0.029090 & 0.098188 & 0.215459 & 0.366585 & 0.503419 \\
    \hline
    \end{tabular}%
  \label{tab:highway-local}%
\end{table}%

\begin{figure}
    \centering

    \subfigure[The process of the infection the nodes in the Germany highway networks \cite{nettt} which are identified by the local structure entropy]{
    \label{High-infect-local1} 
    \centering
    \includegraphics[scale=0.585]{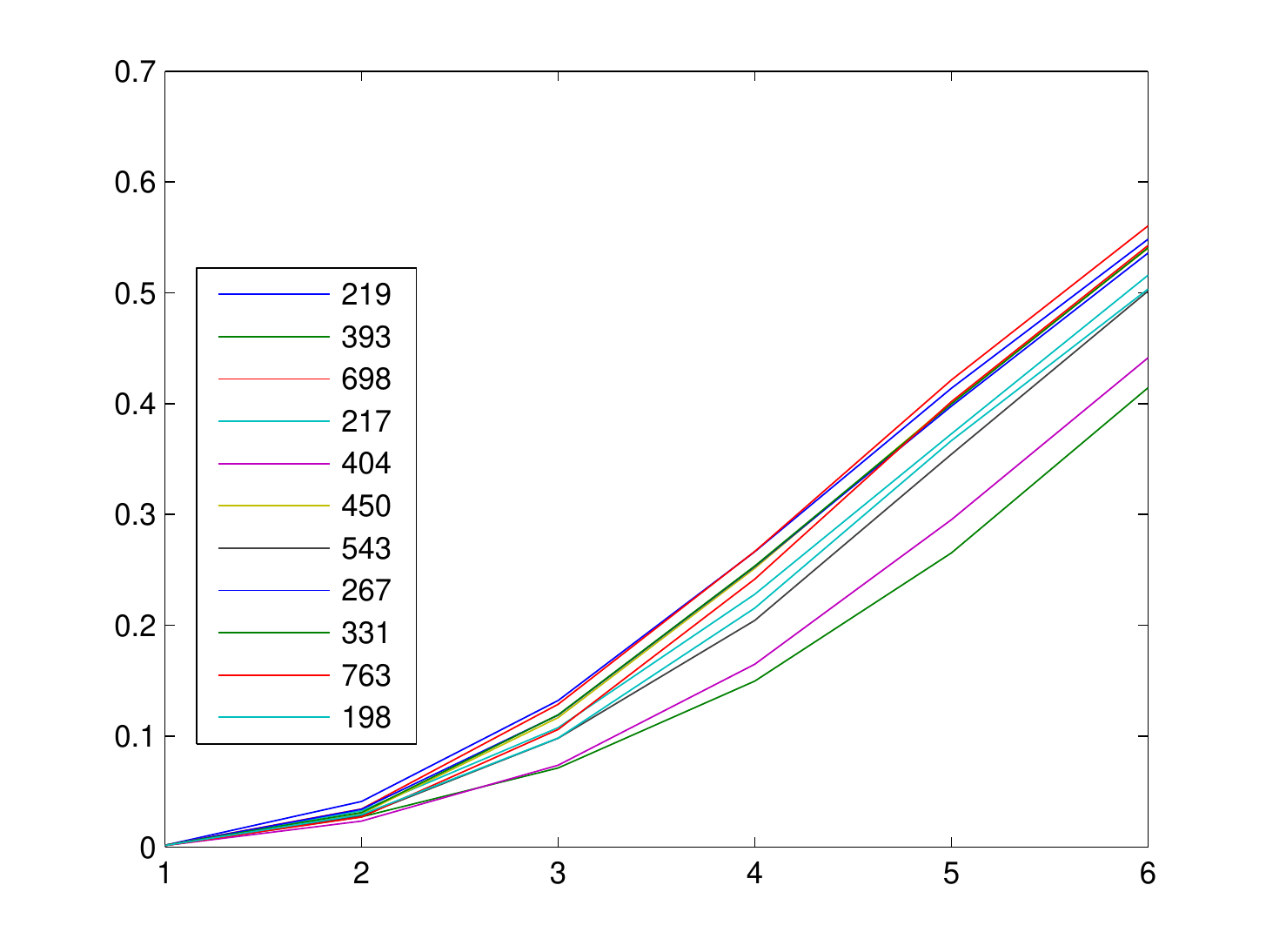}}
  \subfigure[The details in the step 3. ]{
    \label{High-infect-local:a} 
    \centering
    \includegraphics[scale=0.28]{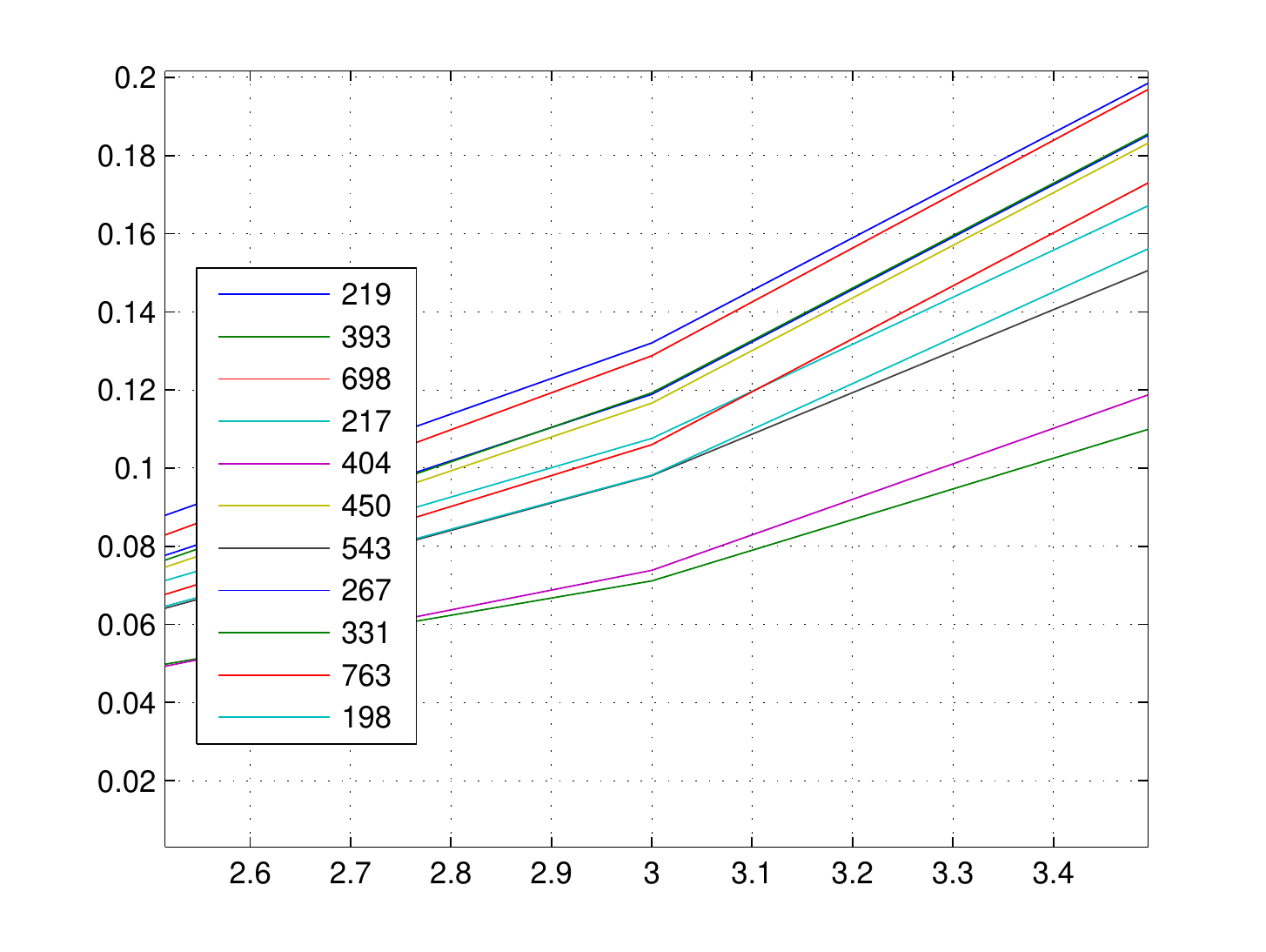}}
    \subfigure[The details in the step 5.]{
    \label{High-infect-local:b} 
    \includegraphics[scale=0.28]{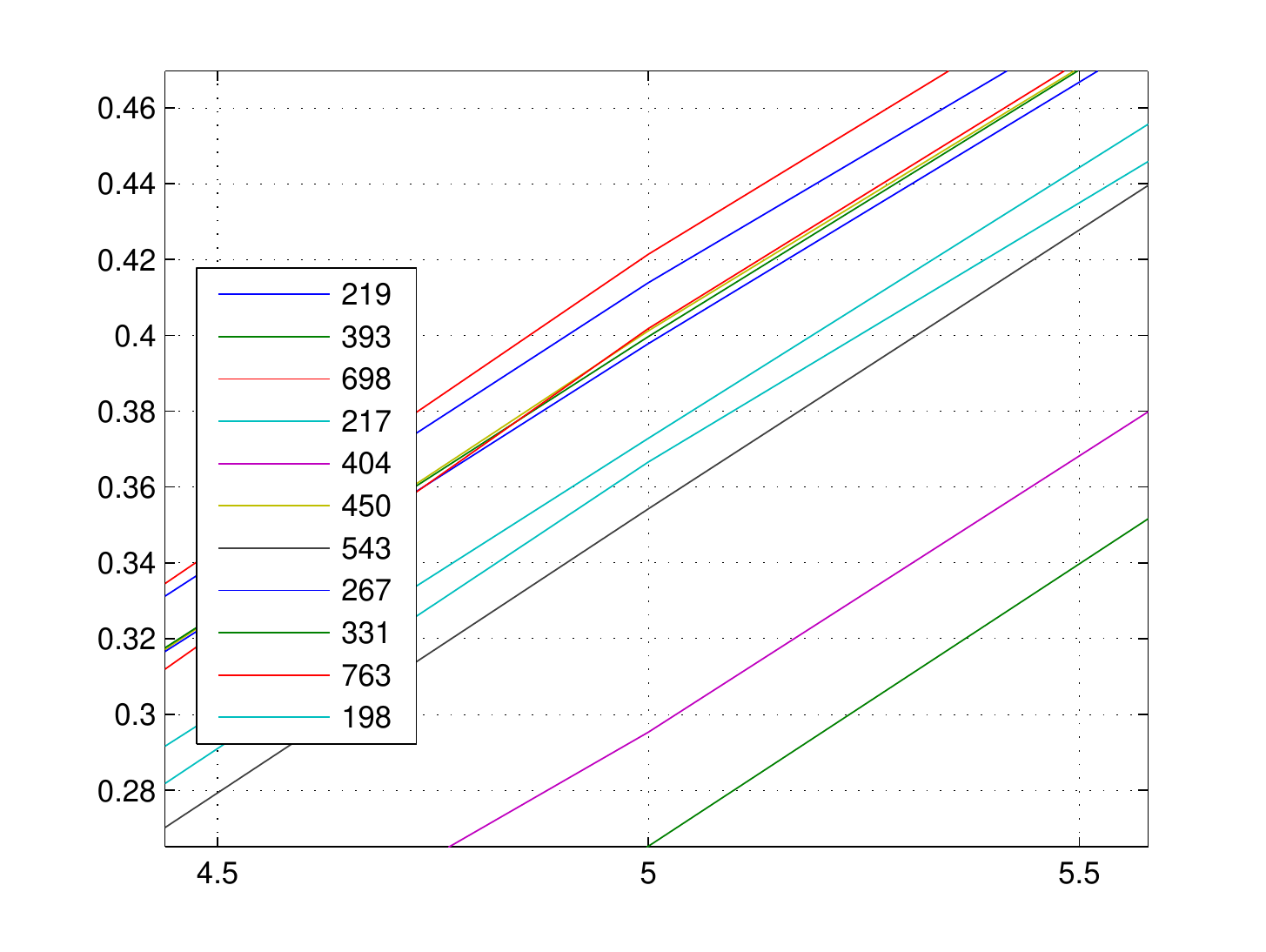}}
    \centering
  \caption{The infected proportion of the most important nodes in the Germany highway networks \cite{nettt} which are identified by the local structure entropy. }\label{High-infect-local}
\end{figure}
The infection processes of the influential nodes in the Germany highway networks \cite{nettt} are shown in the Table \ref{tab:highway-local} and Fig.\ref{High-infect-local}. It is clear that in the Germany highway networks \cite{nettt}, after 100 times infection almost half of the nodes in the network have been infected. The nodes which are identified by the local structure entropy have a big influence in the Germany highway networks \cite{nettt}.

\begin{table}[htpb]
  \centering
  \caption{The infected proportion of the most important nodes in the protein-protein interaction network in budding yeast \cite{networkdata} which are identified by the local structure entropy}
    \begin{tabular}{lrrrrrr}
    \hline
    node  & Step 1 & Step 5 & Step 10 & Step 15 & Step 20 & Step 25 \\
    \hline
    944   & 0.002095 & 0.016147 & 0.073933 & 0.160880 & 0.295533 & 0.443312 \\
    942   & 0.002108 & 0.015949 & 0.074454 & 0.161361 & 0.293070 & 0.443533 \\
    941   & 0.002139 & 0.016166 & 0.074633 & 0.161848 & 0.294371 & 0.444168 \\
    939   & 0.002089 & 0.016160 & 0.074301 & 0.160739 & 0.294250 & 0.443074 \\
    935   & 0.002118 & 0.016304 & 0.074498 & 0.160640 & 0.294867 & 0.441506 \\
    928   & 0.002095 & 0.016043 & 0.075048 & 0.161180 & 0.293736 & 0.441586 \\
    185   & 0.002068 & 0.016388 & 0.074253 & 0.161112 & 0.295613 & 0.443363 \\
    184   & 0.002101 & 0.016072 & 0.074579 & 0.161200 & 0.295501 & 0.442960 \\
    940   & 0.002217 & 0.016855 & 0.077028 & 0.164835 & 0.299863 & 0.449456 \\
    938   & 0.002183 & 0.016952 & 0.077133 & 0.165342 & 0.300889 & 0.448221 \\
    927   & 0.002272 & 0.016595 & 0.077356 & 0.166127 & 0.301810 & 0.448065 \\
    \hline
    \end{tabular}%
  \label{tab:Yeast-local}%
\end{table}%

\begin{figure}
    \centering

    \subfigure[The process of the infection the nodes in the protein-protein interaction network in budding yeast \cite{networkdata} which are identified by the local structure entropy]{
    \label{Yeast-infect-local1} 
    \centering
    \includegraphics[scale=0.585]{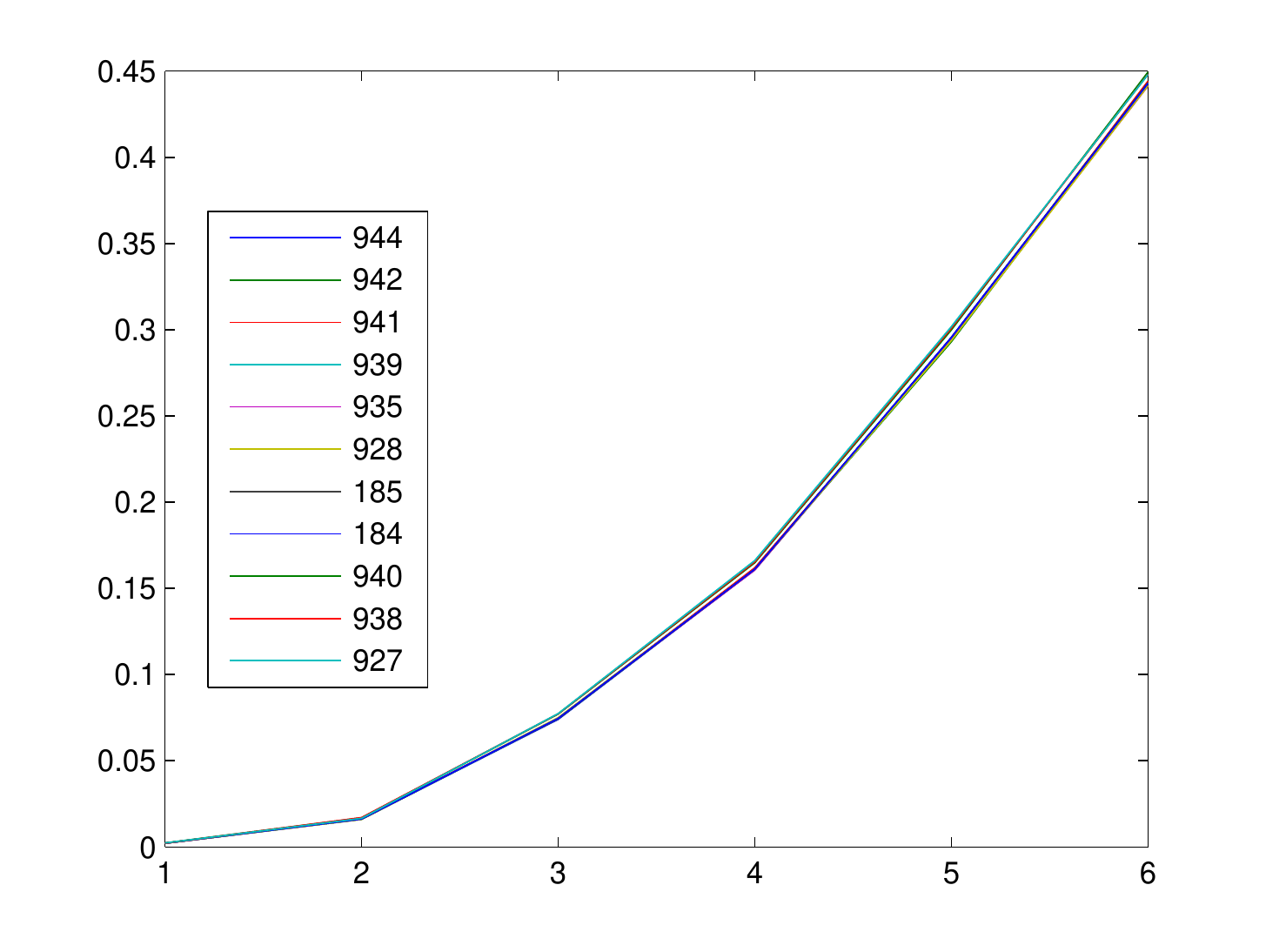}}
  \subfigure[The details in the step 3. ]{
    \label{Yeast-infect-local:a} 
    \centering
    \includegraphics[scale=0.28]{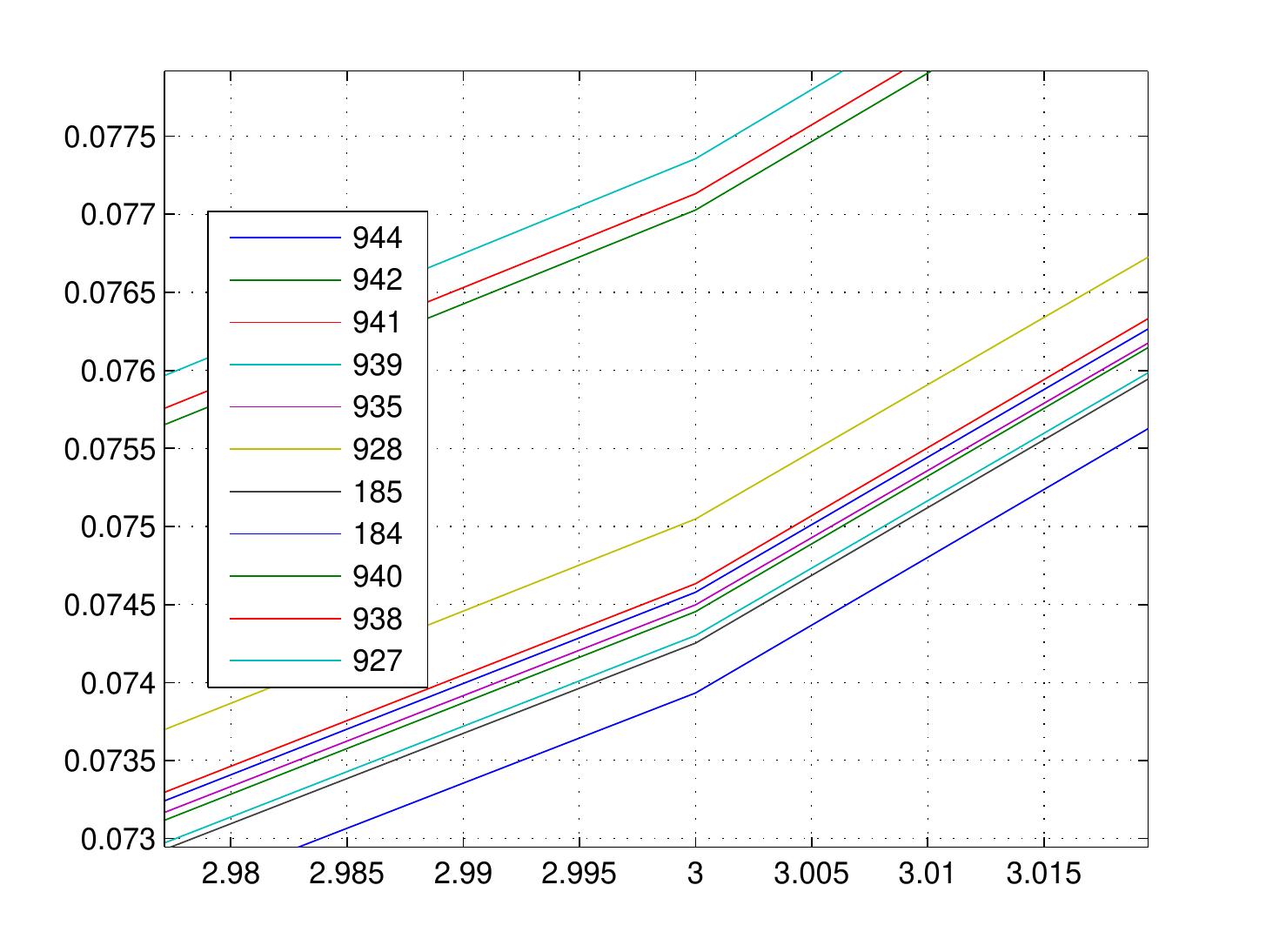}}
    \subfigure[The details in the step 5.]{
    \label{Yeast-infect-local:b} 
    \includegraphics[scale=0.28]{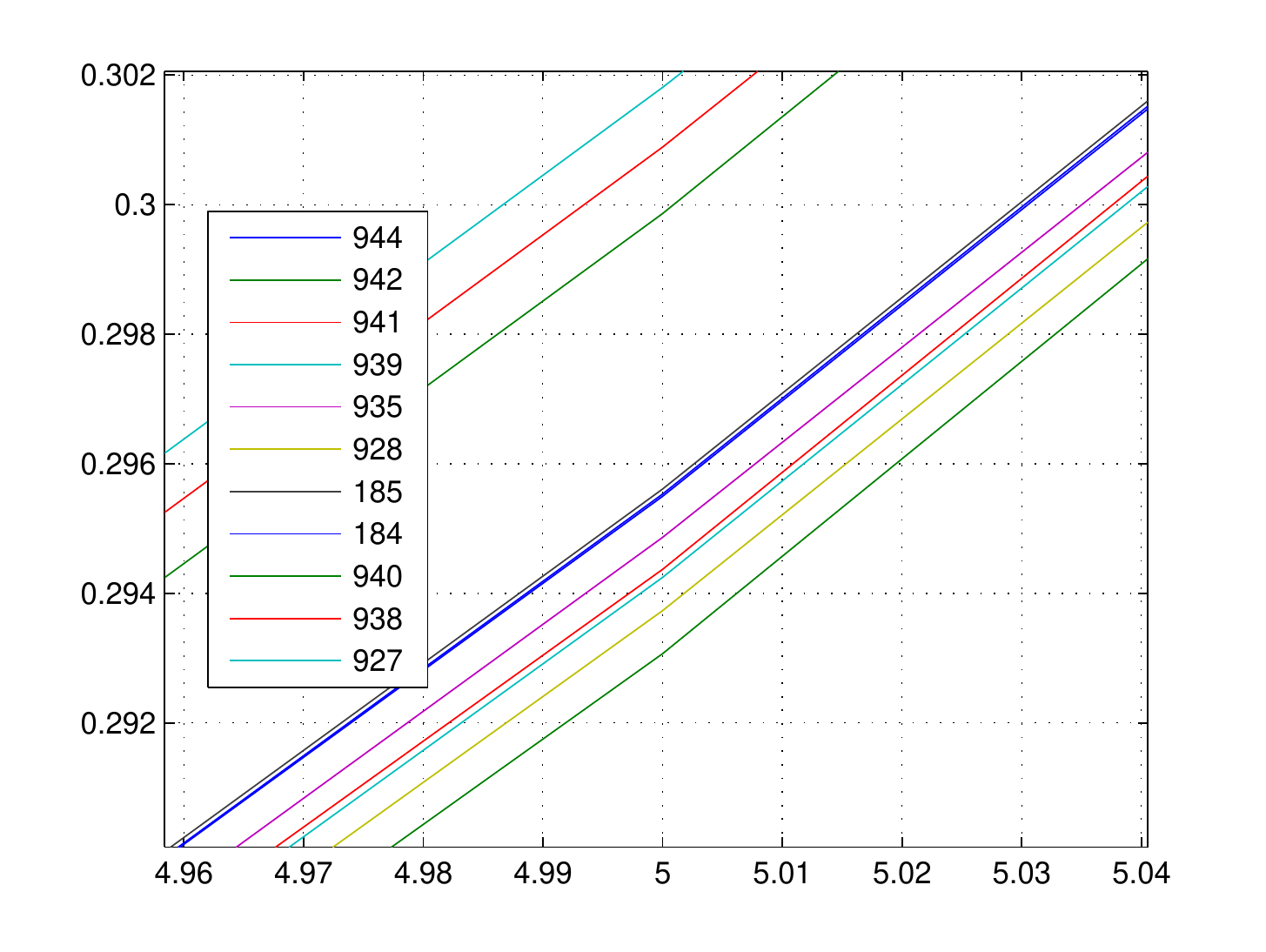}}
    \centering
  \caption{The infected proportion of the most important nodes in the protein-protein interaction network in budding yeast \cite{networkdata} which are identified by the local structure entropy. }\label{Yeast-infect-local}
\end{figure}

The protein-protein interaction network in budding yeast \cite{networkdata} is a biological network. The processes of the infection in the protein-protein interaction network in budding yeast \cite{networkdata} are shown in the Table \ref{tab:Yeast-local} and the Fig.\ref{Yeast-infect-local}. The results show that the nodes which are identified by the local structure entropy have an stable and big influence to the protein-protein interaction network in budding yeast \cite{networkdata}.

The details of our research show that the local structure entropy can identify those nodes which have a small value of degree but have an big influence to the whole network. Most of the nodes which are identified by the local structure entropy are the intermediate connection nodes, they connect those nodes which have a big value of degree.
\section{Conclusion}
\label{conclusion}
Identifying the influential nodes in the network is one of the most important research direction in the research of complex network. It can be used to identify the leader in the social network. Tt can be used to find the central nodes in the power network. It also can be used in the human disease network to find the main gene which control the health of our human. There are many methods can be used to identify the influential nodes in the network from different needs. In this paper, the local structure entropy is proposed based ont the degree centrality and the statistical mechanics. The influence of the local network on the whole network is used to replace the node's influence on the whole networks. In our opinion, the local structure entropy can avoid the complex calculation in the traditional methods and merge the influence of the degree and the betweenness of the nodes in the local network. The results of this paper show that the local structure entropy is efficacious and rationality.
%

\section{Acknowledgment}
The work is partially supported by National Natural Science Foundation of China (Grant No. 61174022), Specialized Research Fund for the Doctoral Program of Higher Education (Grant No. 20131102130002), R$\&$D Program of China (2012BAH07B01), National High Technology Research and Development Program of China (863 Program) (Grant No. 2013AA013801), the open funding project of State Key Laboratory of Virtual Reality Technology and Systems, Beihang University (Grant No.BUAA-VR-14KF-02).
\bibliography{zqreference}

\end{document}